\documentclass[journal=jctcce]{achemso}
\setkeys{acs}{articletitle = true}
\setkeys{acs}{maxauthors = 10}
\setkeys{acs}{etalmode =truncate}
\SectionNumbersOn

\usepackage[english]{babel}
\usepackage{graphicx}
\usepackage{amssymb}
\usepackage{amsmath}
\usepackage{amsfonts}
\usepackage{caption}
\usepackage{subcaption}
\usepackage{siunitx}
\usepackage{color}
\usepackage{bm}
\usepackage[update,prepend]{epstopdf}
\usepackage{multirow}
\usepackage{enumitem}
\usepackage{booktabs}
\newlist{points}{enumerate}{1}
\setlist[points]{label=(\Alph*),ref=(\Alph*)}

\newcommand{\la}{\left\langle}
\newcommand{\ra}{\right\rangle}
\title{Accurate coarse-graining of small organic molecules in melts and thin films using density-dependent potentials}
\author{Sayan Dutta}
\affiliation{Leibniz-Institut f{\"u}r Polymerforschung Dresden e.V., Hohe Stra\ss e 6, 01069 Dresden, Germany}
\alsoaffiliation{Institut f{\"u}r Theoretische Physik, Technische Universit{\"a}t Dresden, 01069 Dresden, Germany}

\author{Maria C. Lesniewski}
\affiliation{Department of Chemistry, Pennsylvania State University, University Park, PA 16802, USA}

\author{Muhammad Nawaz Qaisrani}
\affiliation{Max Planck Institute for Polymer Research, Ackermannweg 10, 55128 Mainz, Germany}

\author{W. G. Noid}
\affiliation{Department of Chemistry, Pennsylvania State University, University Park, PA 16802, USA}

\author{Denis Andrienko}
\affiliation{Max Planck Institute for Polymer Research, Ackermannweg 10, 55128 Mainz, Germany}

\author{Arash Nikoubashman}
\email{anikouba@ipfdd.de}
\affiliation{Leibniz-Institut f{\"u}r Polymerforschung Dresden e.V., Hohe Stra\ss e 6, 01069 Dresden, Germany}
\alsoaffiliation{Institut f{\"u}r Theoretische Physik, Technische Universit{\"a}t Dresden, 01069 Dresden, Germany}

\date{\today}

\begin{document}

\begin{abstract}
Conjugated organic molecules play a central role in a wide range of optoelectronic devices, including organic light-emitting diodes, organic field-effect transistors, and organic solar cells. A major bottleneck in the computational design of these materials is the discrepancy between simulation and experimental time and length scales. Coarse-graining (CG) offers a promising solution to bridge this gap by reducing redundant degrees of freedom and smoothing the potential energy landscape, thereby significantly accelerating molecular dynamics simulations. However, standard CG models are typically parameterized from homogeneous bulk simulations and assume density-independent effective interactions. As a consequence, they often fail to replicate inhomogeneous systems, such as (free-standing) thin films, due to an incorrect representation of liquid-vacuum interfacial properties. In this work, we develop a CG parametrization strategy that incorporates local-density-dependent potentials to capture material heterogeneities. We evaluate the methodology by simulating free-standing films and comparing interfacial orientational order parameters between all-atom and CG simulations. The resulting CG models accurately reproduce bulk densities and radial distribution functions as well as molecular orientations at the liquid–vacuum interface. This work paves the way for reliable, computation-driven predictions of atomically resolved interfacial ordering in organic molecular systems.
\end{abstract}

\maketitle
\section{Introduction}
The high computational cost of all-atom (AA) models has driven significant interest in the development of coarse-grained (CG) models for a wide range of systems, including biomolecular assemblies,\cite{banerjee2024perspective, chen2017data} polymer and soft material design.\cite{zhang2024chemically} By focusing on the most essential features of a system, CG models can achieve substantial computational advantages through their reduced resolution, and also provide conceptual insights into the underlying mechanisms.\cite{deserno2009mesoscopic} However, the removal of atomic-level detail often compromises the transferability and representability of CG models,\cite{louis2002beware, johnson2007representability} limiting their broader applicability. This can manifest as quantitative discrepancies, e.g., artificially accelerated dynamics due to overly smoothed energy landscapes, or even qualitative failures, such as thin-film instabilities arising from inaccurate treatment of many-body effects. These challenges underscore the need for CG models that balance efficiency, fidelity, and transferability. 

Most CG methods can be roughly divided into two categories, top-down approaches that parameterize CG interactions to reproduce macroscopic (thermodynamic) observables, and bottom-up approaches that derive them from atomistic simulations.\cite{delyser2019analysis} In principle, bottom-up models could exactly reproduce the structural and thermodynamic behavior of the underlying AA system,\cite{wagner2016representability, dunn2016van} but in practice CG models often rely on approximations to the true many-body potential of mean force (PMF).\cite{kirkwood1935statistical, akkermans2001structure} A common approximation is to represent intermolecular interactions using pairwise central potentials. In structure-based methods, these potentials are typically tuned to match atomistic radial distribution functions, $g(r)$.\cite{muller2002coarse} This approach can reproduce certain structural features, but often fails to capture higher-order correlations that are important in systems with complex molecular architectures and environments.\cite{rudzinski2012role} Other bottom-up methods, such as force-matching,\cite{izvekov2005multiscale, izvekov2005liquid, noid2008multiscaleI, noid2008multiscaleII} extend beyond $g(r)$ fitting and address some of these limitations, but challenges remain in achieving transferability across thermodynamic conditions.

In particular, conventional CG models often struggle to capture interfacial or inhomogeneous behavior,\cite{jochum2012structure, dalgicdir2013transferable} since their interactions are typically parameterized from homogeneous systems in, e.g., the liquid or vapor phase. A key development for improving the accuracy of CG models was the introduction of one-body local-density dependent potentials (LDPs) in the framework of many-body dissipative particle dynamics,\cite{pagonabarraga2001dissipative, warren2003vapor} inspired by classical density functional theory.\cite{evans1979nature} This framework enables the simulation of, e.g., liquid–vapor equilibria\cite{warren2003vapor, ghoufi2011mesoscale} and phase coexistence in implicit-solvent lipid membrane models.\cite{homberg2010main} Recent bottom-up studies have employed LDPs to construct implicit-solvent models,\cite{allen2008novel, allen2009evaluating, sanyal2016coarse, rosenberger2019transferability} to simulate shock compression of (semi)crystalline polymers,\cite{moore2016coarse, agrawal2016pressure} to determine fluid phase equilibria of binary mixtures,\cite{sanyal2018transferable} and to describe thin films and droplets of small molecules\cite{delyser2017pressure, delyser2020interface, wagner2017extending} and homopolymers.\cite{berressem2021ultra}

Thin films of conjugated organic molecules are a prime application example for CG models with LDPs: These molecules are crucial building blocks for various optoelectronic applications, including organic light-emitting diodes, \cite{hong2021OS, shinar2013OS} field-effect transistors,\cite{liu2022FET} and solar cells.\cite{huang2022SC} Their appeal lies in the combination of tunable electronic structure, mechanical flexibility, and solution processability afforded by organic synthesis. In practice, these molecules function as active layers in thin-film architectures, where device performance depends sensitively on both large-scale film stability and the local orientation of individual molecules at interfaces. Exploring the vast chemical design space remains a long-standing challenge, as optimizing composition and morphology is experimentally intensive and structure–property relationships are often non-trivial.\cite{yaghi2003retsys, fried2001design} Simulations can accelerate this process, provided that the simulations are both efficient and accurate. For example, predicting energetic disorder and charge carrier mobilities has proven feasible in systems that are either fully amorphous or perfectly crystalline.\cite{mondal2021molecular, sachnik2023elimination, scherer2024predicting} However, capturing (non-equilibrium) morphologies that emerge under processing conditions, where partial ordering, interfacial effects, and local heterogeneity are key, remains difficult. 

\begin{figure}[ht]
    \centering
    \includegraphics[width=10.0cm]{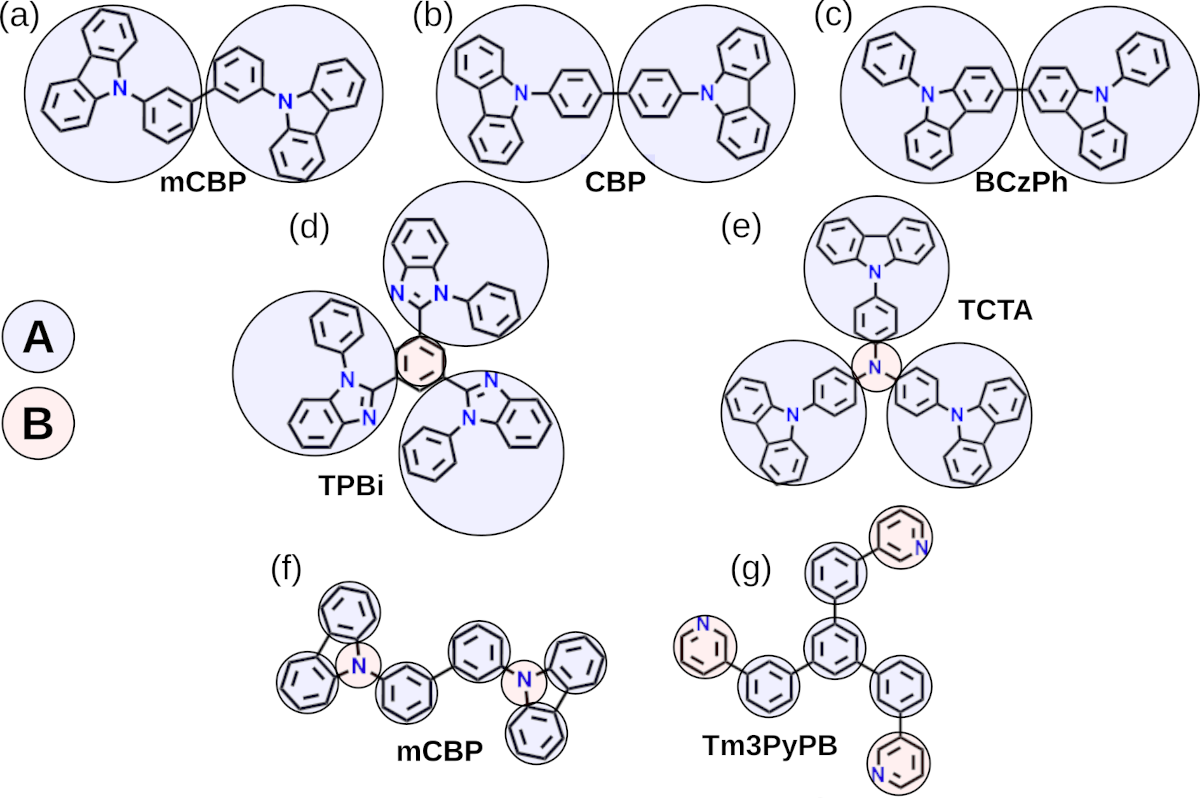}
    \caption{Chemical structures of molecules with CG mapping schemes. (a-c) Low resolution two-site models, (d,e) Low resolution four-site models, and (f,g) high resolution multi-site models.}
    \label{fig:structures}
\end{figure}

In this study, we demonstrate that CG models augmented with LDPs, which have been parametrized on atomistic simulations, are capable of accurately predicting molecular alignment in thin films across a broad range of small conjugated molecular species (Fig.~\ref{fig:structures}). We present the CG procedure in Sec.~\ref{sec:model.CG}, and discuss the results for organic molecules in bulk melts and in thin films in Secs.~\ref{sec:results.bulk} and \ref{sec:results.film}, respectively. Finally, the computational efficiency of our CG model is evaluated in Sec.~\ref{sec:results.speedup}.

\section{Coarse-Graining Procedure}
\label{sec:model.CG}
As the first step, we map the AA configuration of $n$ atoms ($\mathbf{r} = \left\{\mathbf{r}_1,\dots,\mathbf{r}_n\right\}$) to a CG configuration consisting of $N < n$ sites ($\mathbf{R} = \left\{\mathbf{R}_1,\dots,\mathbf{R}_N\right\}$). We have considered different mapping schemes for the molecules of interest, as illustrated in Fig.~\ref{fig:structures}. Although the investigated molecules are rather small, there can be different possible mapping schemes. Based on the molecular architecture, we have separated them in three systematic different categories: (i) Low resolution mapping to A-A dimers for mCBP, CBP, and BCzPh, which share similar mirror symmetry. (ii) Low resolution mapping to four-site models with two particle types for TPBi and TCTA. In TPBi, each benzimidazole unit together with its attached terminal phenyl is grouped into a single CG site to capture the overall rigidity and extended conjugation of the fused system. The central phenyl ring connecting the three arms is treated as a separate particle type due to its distinct local environment and connectivity. TCTA exhibits a similar structural motif, i.e., planar rigid carbazole-phenyl groups (as in mCBP, CBP, and BCzPh) connected through a central nitrogen atom. (iii) High resolution mapping with two particle types for Tm3PyPB to retain its chemically distinct heterocyclic units. To test the effect of mapping scheme, we have also used such a higher resolution CG representation for the molecule mCBP. 

Formally, the CG representations are created using the operator $\mathbf{M}(\mathbf{r})$, which determines the canonical mapped distribution
\begin{equation}
    p(\mathbf{R}) = \int \text{d}\mathbf{r} \, p(\mathbf{r}) \, \delta\left[\mathbf{R} - \mathbf{M}(\mathbf{r})\right],
\end{equation}
where $p(\mathbf{r})$ is the AA probability distribution. We seek to approximate the many-body potential of mean force (PMF) $W$,\cite{noid2013systematic, likos2001effective} which we define by the total Boltzmann weight of AA configuration $\textbf{r}$ mapped to the corresponding CG configuration $\textbf{R}$ 
\begin{equation}
    \exp[-\beta W(\mathbf{R})] \propto z_{\mathbf{R}}(\mathbf{R}) \, V^{n - N},
\end{equation}
where $z_{\mathbf{R}}(\mathbf{R})$ is the restricted configuration integral, defined as 
\begin{equation}
z_{\mathbf{R}}(\mathbf{R}) = \int \text{d}\mathbf{r} \, \exp[-\beta U(\mathbf{r})] \, \delta\left[\mathbf{R} - \mathbf{M}(\mathbf{r})\right] .
\end{equation}
If we can get the PMF as the function of both configuration and also the thermodynamic state point, then the CG model can perfectly reproduce the mapped configuration distribution at different state points. Calculation of thermodynamic properties heavily depend on this state point dependence.\cite{dunn2016van} 

\begin{figure*}[ht]
\centering
\includegraphics[width=\linewidth]{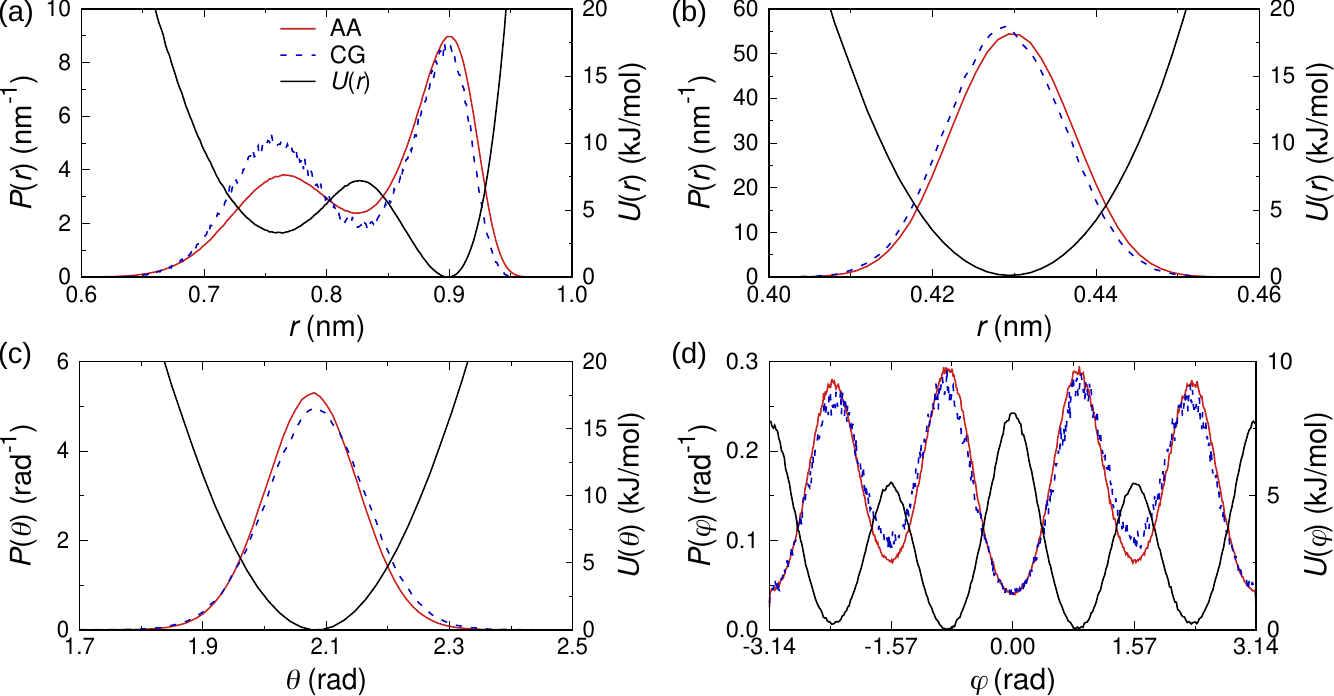}
\caption{Comparison of probability distribution $P$ of bonded particles from AA mapped (solid red lines) and CG simulations (dashed blue lines) for (a) mCBP and (b-d) Tm3PyPB. The black lines show the bonded interactions used in the CG simulations. All data computed at $T=550\,\text{K}$.}
\label{fig:Ub}
\end{figure*}

We decompose the total potential
\begin{equation}
    U(\mathbf{R}) = U_\text{b}(\mathbf{R}) + U_\text{nb}(\mathbf{R}) ,
\end{equation} 
where $U_\text{b}$ includes intramolecular contributions that govern bond, angle, and torsional degrees of freedom, while $U_\text{nb}$ describes non-bonded interactions that are modeled by both pair and local-density-dependent potentials:
\begin{equation}
    U_\text{nb}(\mathbf{R}) = U_2(\mathbf{R}) + U_\rho(\mathbf{R}) .
\end{equation}
We have determined $U_\text{b}$ through Boltzmann inversion of the mapped AA trajectory, as shown in Fig.~\ref{fig:Ub} for selected cases. Typically, higher resolution models require a larger set of bonded interactions, such as multiple bond types, angle bending, and torsional dihedrals, to preserve the target conformations more accurately.

To determine the non-bonded interactions $U_\text{nb}$, we first removed all bonded forces from the AA trajectory\cite{ruhle2011hybrid}, and assigned a type, $t_I$, to each CG site $I$. We define $U_{(\tau,\tau')}(r)=U_{(\tau',\tau)}(r)$ as the central pair potential governing pair non-bonded interactions between particles of types $\tau$ and $\tau'$. The pair contribution to $U_\text{nb}$ is defined as
\begin{equation}
    U_2(\mathbf{R}) = \sum_{(I,J)} U_{(t_I,t_J)}(R_{IJ}) ,
\end{equation}
where $(I,J)$ identifies a pair of particles and $R_{IJ}$ is the corresponding pair distance. This sum contains all pairs that interact via non-bonded interactions and, in particular, reflects the excluded interactions that are determined by the topology of the CG molecules. In the present work, we exclude all non-bonded interactions from intramolecular pairs that are separated by fewer than 4 bonds.

In order to define the LD contribution, $U_\rho,$ we first define the local density of $\tau$ sites around particle $I$ as a sum of weighted contributions from neighboring particles\cite{pagonabarraga2001dissipative}
\begin{equation}
\label{eq:def-rhotI}
    \rho_{\tau|I} = \sum_{J} w_{\tau|t_I}(R_{IJ})	.
\end{equation}
Specifically, this sum goes over all CG sites $J$ that are of type $\tau$ and that interact with site $I$ through a non-bonded pair interaction. The weighting function, $w_{\tau|t_I}$, in Eq.~\eqref{eq:def-rhotI} depends upon both $\tau$ and also the type, $t_I$, of the particle $I$. In the following we adopt the (scaled) Lucy function\cite{lucy1977numerical, wagner2017extending}
\begin{equation}
    \label{eq:def-wttp}
    w_{\tau|\tau'}(r) = 
		[w_{\tau|\tau'}]^{-1}
		\left(1 - \frac{r}{L_{\tau|\tau'}}\right)^3 \left(1 + \frac{3r}{L_{\tau|\tau'}}\right) \theta\left(1 - \frac{r}{L_{\tau|\tau'}}\right)	,
\end{equation}
where $\theta(x)$ is the Heaviside function and $[w_{\tau|\tau'}] = 16\pi L_{\tau|\tau'}^3/105$ is a normalization such that $\int_0^\infty \text{d} r \:  4\pi r^2 w_{\tau|\tau'}(r) = 1$. Equation~\eqref{eq:def-wttp} ensures that $w_{\tau|\tau'}$ monotonically decreases from $r = 0$ and vanishes beyond the LD length-scale, $L_{\tau|\tau'}$. Finally, the LD contribution, $U_\rho$, to the total potential is defined by
\begin{equation}
    \label{eq:def-Uld}
    U_\rho(\mathbf{R}) = \sum_I \sum_{\tau} U_{\tau|t_I}(\rho_{\tau|I})	,
\end{equation}
where $U_{\tau|\tau'}$ is an LDP that depends on the density of $\tau$ sites around a fixed particle of type $\tau'$.

At this point, it is useful to make several comments about the LDP. In general, the LD length-scales, weighting functions, and potentials are not necessarily symmetric with respect to site types, i.e., $L_{\tau|\tau'} \neq L_{\tau'|\tau}$, $w_{\tau|\tau'} \neq w_{\tau'|\tau}$, and $U_{\tau|\tau'} \neq U_{\tau'|\tau}$.\cite{sanyal2018transferable, delyser2019analysis} In particular, it is sometimes convenient to treat a finite LDP to govern the density of $\tau$ sites around central $\tau'$ sites, $U_{\tau|\tau'}(\rho) \neq 0$, while ignoring the complementary LDP, $U_{\tau'|\tau}(\rho) = 0$. Additionally, note that the pair and LD interactions reflect the same bonded exclusions. Consequently, the LDP $U_{\tau|\tau'}$ can employ the neighbor list for the corresponding pair potential $U_{(\tau,\tau')}$, as long as the the LD length scale, $L_{\tau|\tau'}$, is not longer than the cut-off defined for $U_{(\tau,\tau')}$. Finally, in some simulations we have supplemented Eq.~\eqref{eq:def-rhotI} with a ``self-term'' for $\tau = t_I$ that corresponds to including particle $I$ in the definition of $\rho_{\tau|I}$.\cite{trofimov2002thermodynamic, delyser2019analysis} This self-term shifts the definition of $\rho_{\tau|I}$ for $\tau = t_I$ by $[w_{\tau\tau}]^{-1}$ but has no impact upon the simulated forces or resulting equilibrium distribution.

Because Eq.~\eqref{eq:def-rhotI} defines the local density around each particle by a sum of distance-dependent contributions from neighboring particles, the local density potential is short-ranged and generates pair-additive forces along the vector between each pair of particles.\cite{pagonabarraga2001dissipative}
In particular, the total non-bonded force on particle $I$ may be expressed:
\begin{equation}
    \label{eq:def-bFInb}
    \mathbf{F}_I^\text{nb} \equiv -\nabla_I U_\text{nb}(\mathbf{R}) =  \sum_{J} \mathbf{F}^\text{nb}_{IJ}(\mathbf{R}) ,
\end{equation}
where the summation goes over all particles $J$ that interact with $I$ through non-bonded interactions. Moreover, the total non-bonded force from particle $J$ may be expressed
\begin{equation}
    \label{eq:def-FIJnb}
    \mathbf{F}^\text{nb}_{IJ}(\mathbf{R})  = F_{(I,J)}^\text{nb}(\mathbf{R}) \widehat{\mathbf{R}}_{IJ}
\end{equation}
where $\widehat{\mathbf{R}}_{IJ}$ is the unit vector pointing from particle $J$ to particle, $I$. If we denote the types of sites $I$ and $J$ by $t_I = \tau$ and $t_J = \tau'$, then the  magnitude of this total non-bonded pair force is\cite{delyser2019analysis}
\begin{equation}
\label{eq:def-FIJnb-magnitude}
F_{(I,J)}^\text{nb}(\mathbf{R}) 
	= 
		F_{(\tau,\tau')}(R_{IJ}) 
	+ 	
		F_{\tau'|\tau}(\rho_{\tau'|I}) w'_{\tau'|\tau}(R_{IJ})
	+
		F_{\tau|\tau'}(\rho_{\tau|J}) w'_{\tau|\tau'}(R_{IJ})
	=
		F_{(J,I)}^\text{nb}(\mathbf{R}) 
		,
\end{equation}
where $F_{(\tau,\tau')}(r) = - \text{d} U_{(\tau,\tau')}(r)/\text{d}r$, $F_{\tau'|\tau}(\rho) = - \text{d} U_{\tau'|\tau}(\rho)/\text{d}\rho$, and $w'_{\tau'|\tau}(r) = - \text{d} w_{\tau'|\tau}(r)/\text{d}r$. Thus, we see that $\mathbf{F}^\text{nb}_{IJ}(\mathbf{R}) = -\mathbf{F}^\text{nb}_{JI}(\mathbf{R})$. Importantly, because they generate short-ranged pair-additive forces, LDPs provide the same computational scaling as conventional pair potentials.\cite{pagonabarraga2001dissipative, sanyal2016coarse}

We determine the non-bonded potential, $U_\text{nb}$, according to a hybrid\cite{ruhle2011hybrid} force-matching variational principle\cite{ercolessi1994interatomic, chorin2003conditional, izvekov2005liquid, izvekov2005multiscale} after eliminating the bonded forces from the AA trajectory, i.e., we minimized
\begin{equation}
\label{eq:def-FM}
\chi_\text{nb}^2[U_\text{nb}] 
	= 
		\frac{1}{3N}
		\la 
					\sum_{I=1}^{3N}
				\left|
					\mathbf{F}_I^\text{nb}(\mathbf{M}(\mathbf{r})) - \mathbf{f}_I^\text{nb}(\mathbf{r}) 
				\right|^2
			\ra	,
\end{equation}
where $\mathbf{F}_I^\text{nb}$ is defined by Eq.~\eqref{eq:def-bFInb}, $\mathbf{f}_I^\text{nb}(\mathbf{r})$ is the net non-bonded force on particle $I$ in the AA configuration, and the angular brackets denote an average over the canonical ensemble for the AA model. 
Specifically, we represented each pair and LD force function on a uniformly spaced grid with cubic splines and solved the corresponding normal system of equations\cite{noid2007multiscale, noid2008multiscaleI}
 with the bottom-up open-source coarse-graining software (BOCS) package.\cite{dunn2017bocs, mjlbocs}

We note that Eq.~\eqref{eq:def-FIJnb-magnitude} implies that the minimizing pair and LD potentials are not unique. Specifically, for each pair of site types $\tau$ and $\tau'$, there exists a non-trivial family of pair and LD potentials that generate the same forces.\cite{delyser2019analysis} In order to ensure the stability of the resulting simulations, we extrapolated the resulting pair and LD potentials into regimes that were not adequately sampled in the AA simulations. In most cases, we employed the protocol of Ref.~\citenum{szukalo2023temperature} to extrapolate each LDP to very low and very high local densities based upon Boltzmann inversion of the corresponding LD distribution. In rare cases that the LD distribution deviates significantly from Gaussian, we extrapolated the LD force functions based upon a quadratic fit to the calculated force functions.

CG simulations were performed in LAMMPS (27 Oct, 2021)\cite{LAMMPS_plimpton1995fast, LAMMPS_thompson2022lammps}, modified to support LD potentials.\cite{delyser2019analysis} Temperature was regulated with a Nos{\'e}–Hoover thermostat\cite{nose2002molecular, hoover1985canonical} (\SI{100}{\femto\second} damping, chain length $n=3$) and pressure with an MTTK barostat\cite{martyna1992nose} (\SI{1}{\pico\second} damping). Simulations included a \SI{1}{\nano\second} equilibration period followed by a \SI{10}{\nano\second} production runs, with a \SI{2}{\femto\second} timestep.

\begin{table}[ht]
\centering
\renewcommand{\arraystretch}{1.3} 
\begin{tabular}{|p{6cm}|p{7cm}|}
\hline
\textbf{Systems} & \textbf{CG non-bonded interactions} \\
\hline
mCBP (2-site), CBP, BCzPh & Pair A--A + LD A$|$A\\
\hline
TPBi, TCTA, mCBP (8-site) & Pair A--A, A--B, B--B + LD A$|$A\\
\hline
Tm3PyPB & Pair A--A, A--B, B--B + LD A$|$A, A$|$B, B$|$A\\
\hline
\end{tabular}
\caption{Different non-bonded interactions based on mapping categories and molecular systems.}
\label{tab:interactions}
\end{table}

\section{Results and Discussion}
\subsection{Bulk properties}
\label{sec:results.bulk}
We considered six different organic molecules with diverse mapping schemes, as shown in Fig.~\ref{fig:structures}. We performed AA simulations as a reference framework for developing CG force fields and evaluating their accuracy, ensuring that structural and thermodynamic properties are faithfully retained. AA simulations were performed in GROMACS 2019.6,\cite{abraham2015GROMACS, pronk2013GROMACS} as described in Sec.~\ref{sec:methods} and the SI.

\begin{figure*}[htb]
    \centering
    \includegraphics[width=\linewidth]{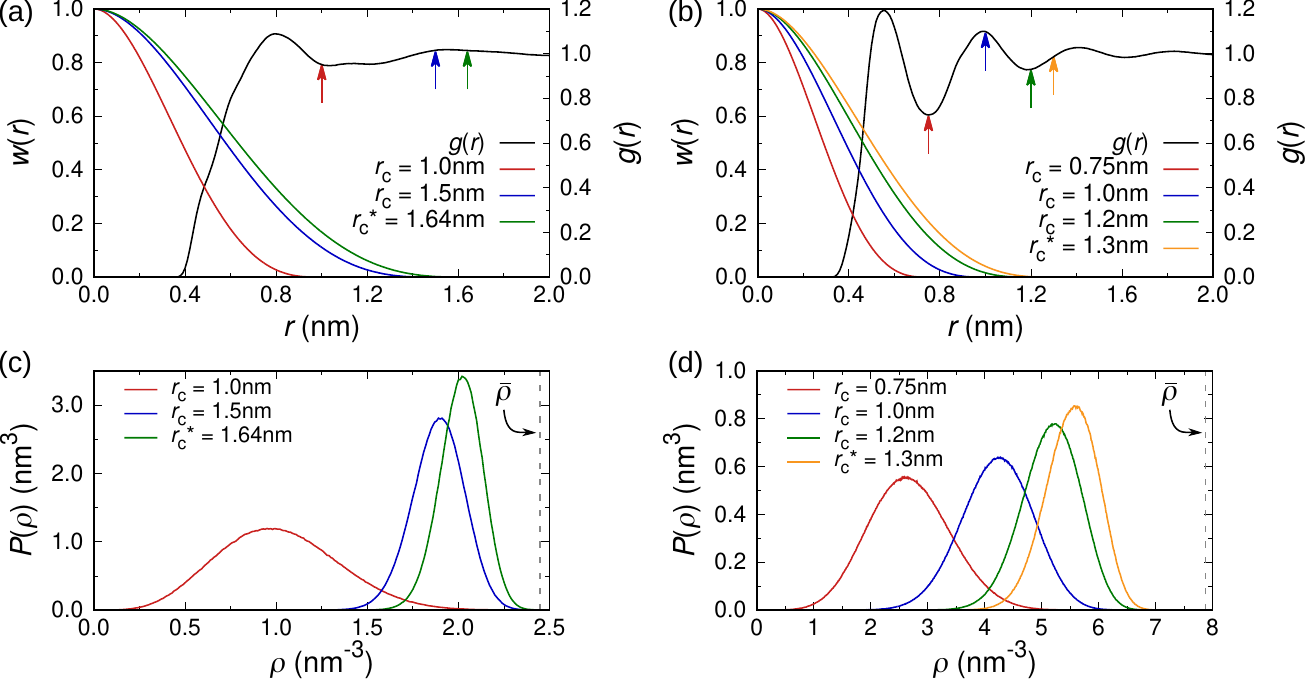}
    \caption{(a,b) Radial distribution functions $g(r)$ (right $y$-axes) and weight functions $w(r)$ (left $y$-axes) from the AA simulations of mCBP, using the (a) two-site and (b) eight-site mapping for various cutoff radii $r_\text{c}$, as indicated. (c,d) Local-density distributions $P(\rho)$ corresponding to the choices of $w(r)$ shown in (a,b). The vertical dashed lines indicate the mean number densities $\bar{\rho} \approx 2.45\,\text{nm}^{-3}$ and $\approx 7.88\,\text{nm}^{-3}$ of the two- and eight-site model, respectively. All simulations performed in bulk at $T=550\,\text{K}$.}
    \label{fig:RDF_wr}
\end{figure*}

To illustrate the CG procedure for determining the LDP, $U_\rho$, we first focus on the example of mCBP in bulk under $NPT$ conditions. As discussed in Sec.~\ref{sec:model.CG}, the choice of cutoff radius $r_\text{c}$ in the weight function $w(r)$ strongly influences the contribution of neighboring sites. To estimate a reasonable range for $r_\text{c}$, we computed the intermolecular radial distribution function $g(r)$ of the CG sites from the mapped AA simulations [Fig.~\ref{fig:RDF_wr}(a,b)]. The two-site model of mCBP exhibits a maximum near $r \approx 0.8\,\text{nm}$, followed by a much weaker second peak around $r \approx 1.5\,\text{nm}$. The higher resolution CG model exhibits a more structured $g(r)$, as expected, with a maximum at $r \approx 0.55\,\text{nm}$, corresponding to van der Waals contacts between two phenyl rings. This maximum is followed by several, increasingly weaker peaks at $r \approx 1.0\,\text{nm}$ and $r \approx 1.4\,\text{nm}$. Guided by these characteristic distances, we then computed the local-density distribution $P(\rho)$ around the mapped sites for different values of $r_\text{c}$, finding a distinct sharpening of $P(\rho)$ with increasing $r_\text{c}$ [Fig.~\ref{fig:RDF_wr}(c,d)]. This trend reflects the fact that, as the sampling volume grows with $r_\text{c}$, the local density approaches the global average density, and $P(\rho)$ narrows toward the (very sharp) global density distribution.\cite{szukalo2023temperature}

\begin{figure*}[htb]
    \centering
    \includegraphics[width=\linewidth]{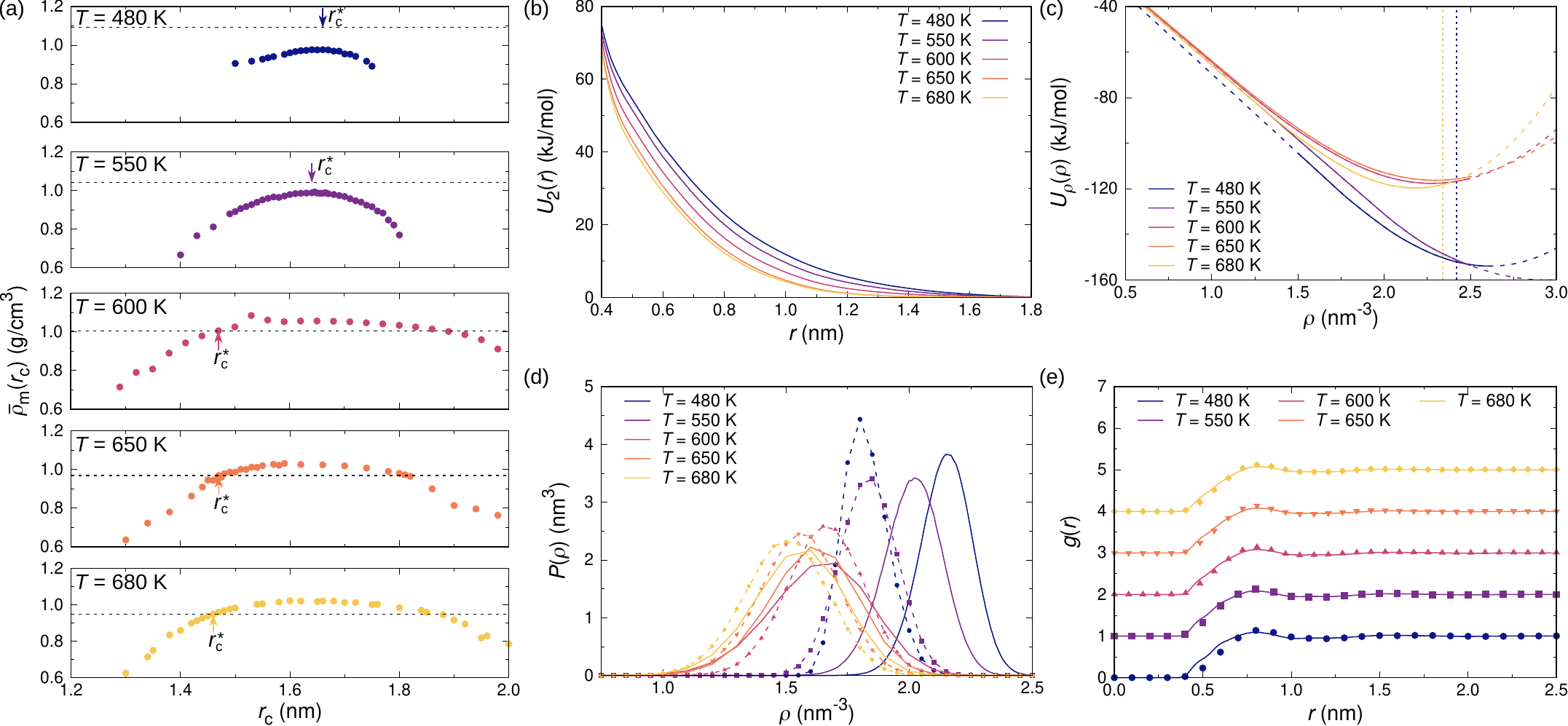}
    \caption{(a) Average mass density $\bar{\rho}_\text{m}$ at different temperatures $T$ as functions of cutoff radius $r_\text{c}$. The horizontal dotted line represents the reference AA mass density, $\bar{\rho}^\text{AA}_\text{m}$. The arrows indicate the chosen optimum cutoff radius $r_\text{c}^*$. (b) Pair potential $U_2(r)$ and (c) LDP $U_\rho$ at different $T$, optimized at $r_\text{c}^*$. The dashed lines indicate the extrapolated $\rho$-region of $U_\rho$. The two vertical lines indicate $\bar{\rho}^\text{CG}_\text{m}$ from CG simulations at $T=480\,\text{K}$ and $680\,\text{K}$, respectively. (d) Local-density distributions $P(\rho)$ from mapped AA simulations (solid lines) and from CG simulations (dashed lines with symbols). (e) Intermolecular radial distribution function $g(r)$ from mapped AA simulations (solid lines) and from CG simulations (symbols). All data for two-site mCBP model.}
    \label{fig:mCBP}
\end{figure*}

Next, we have applied the force-matching variational principle to determine the effective potential $U = \{U_2, U_\rho\}$ for different choices of $r_\text{c}$. To highlight the effect of the density-dependent term, we have shifted $U_2$ and $U_\rho$ such that $U_2$ matched the effective pair potential $U_2'$ \textit{without} local density contributions (Fig.~S2 in the SI). We then performed a series of CG simulations, parameterized at different $r_\text{c}$, to determine the average mass density $\bar{\rho}_\text{m}$. Each system was simulated for \SI{2}{\nano\second}--\SI{5}{\nano\second}, which was sufficient to reach equilibrium. Figure~\ref{fig:mCBP} compares the mass densities from the CG ($\bar{\rho}_\text{m}^\text{CG}$) and AA ($\bar{\rho}_\text{m}^\text{AA}$) simulations at different $T$, revealing several important trends: With increasing $r_\text{c}$, $\bar{\rho}_\text{m}^\text{CG}$ initially rises, goes through a rather shallow maximum -- slightly above $\bar{\rho}_\text{m}^\text{AA}$ for all investigated temperatures $T \geq 600\,\text{K}$ -- and then decreases again. Furthermore, the optimum cutoff radius $r_\text{c}^*$, defined by $\bar{\rho}_\text{m}^\text{CG} \approx \bar{\rho}_\text{m}^\text{AA}$, decreases with increasing $T$. Notably, at most temperatures, $r_\text{c}^*$ is close to the value of $r_\text{c}$, where the force-matching error $\chi_\text{nb}^2$ [see Eq.~\eqref{eq:def-FM}] is minimized [Fig.~S3 in the SI]. These trends indicate a (weak) temperature dependence of the LDP, which we will explore in more detail below.

Examining the effective interaction potentials $U_2$ and $U_\rho$ in Figs.~\ref{fig:mCBP}(b) and (c), respectively, we find that $U_2$ is purely repulsive and primarily captures excluded volume interactions between CG sites. In contrast, the LDP is strongly attractive at intermediate number densities, providing the cohesive interactions necessary to stabilize the melt under the investigated $NPT$ conditions. At number densities above the average density, $\rho \gtrsim \bar{\rho} \approx 2.5\,\text{nm}^{-3}$, $U_\rho$ becomes increasingly repulsive, consistent with the finite compressibility of the system. The combination of these two terms, $U = \{U_2, U_\rho\}$, thus balances short-range excluded volume with density-dependent cohesion and repulsion. 

For $T \geq 600\,\text{K}$, the CG simulations also capture the local-density distribution, $P(\rho)$, of the mapped AA simulations rather accurately [Fig.~\ref{fig:mCBP}(d)]. For the two lower temperatures $T=480\,\text{K}$ and $550\,\text{K}$, however, the $P(\rho)$ from the CG simulations are shifted to considerably lower number densities, consistent with the reduction in mass density, $\bar{\rho}^\text{CG}_\text{m} < \bar{\rho}^\text{AA}_\text{m}$, observed in these cases [Fig.~\ref{fig:mCBP}(a)]. Importantly, CG simulations with these potentials still reproduce the pair correlation of the reference AA simulations, as demonstrated by the excellent agreement of the intermolecular $g(r)$ data [Fig.~\ref{fig:mCBP}(e)].

We now turn to the higher resolution model of mCBP, in which each phenyl ring is represented by a CG bead of type A, while the two nitrogen atoms are retained as distinct beads of type B [Fig.~\ref{fig:structures}(f)]. In this representation, each B bead is embedded within a local environment formed by three slightly larger A beads, so that their positions are strongly correlated and the B sites are effectively buried within the density cloud of the surrounding A beads. Consequently, local-density fluctuations experienced by B beads are largely governed by the arrangement of A beads. To simplify the parametrization procedure, we therefore introduce an LDP only for the A beads, while keeping the B-related interactions density independent. Repeating the same systematic scan of $r_\text{c}$ for the higher resolution model, we identified an optimal cutoff $r_\text{c}^* = 1.3\,\text{nm}$ at $T=550\,\text{K}$, which is close to the third peak in $g(r)$ for the mapped A-A pair correlation [cf. Fig.~\ref{fig:RDF_wr}(b)]. This distance is substantially shorter than the corresponding $r_\text{c}^* \approx 1.65\,\text{nm}$ obtained for the low resolution model. This decrease in $r_\text{c}^*$ presumably reflects the smaller bead sizes (and larger average number densities $\bar{\rho}$) in the high resolution model, which allow the particles to pack more tightly.

\begin{figure*}[htb]
    \centering
    \includegraphics[width=\linewidth]{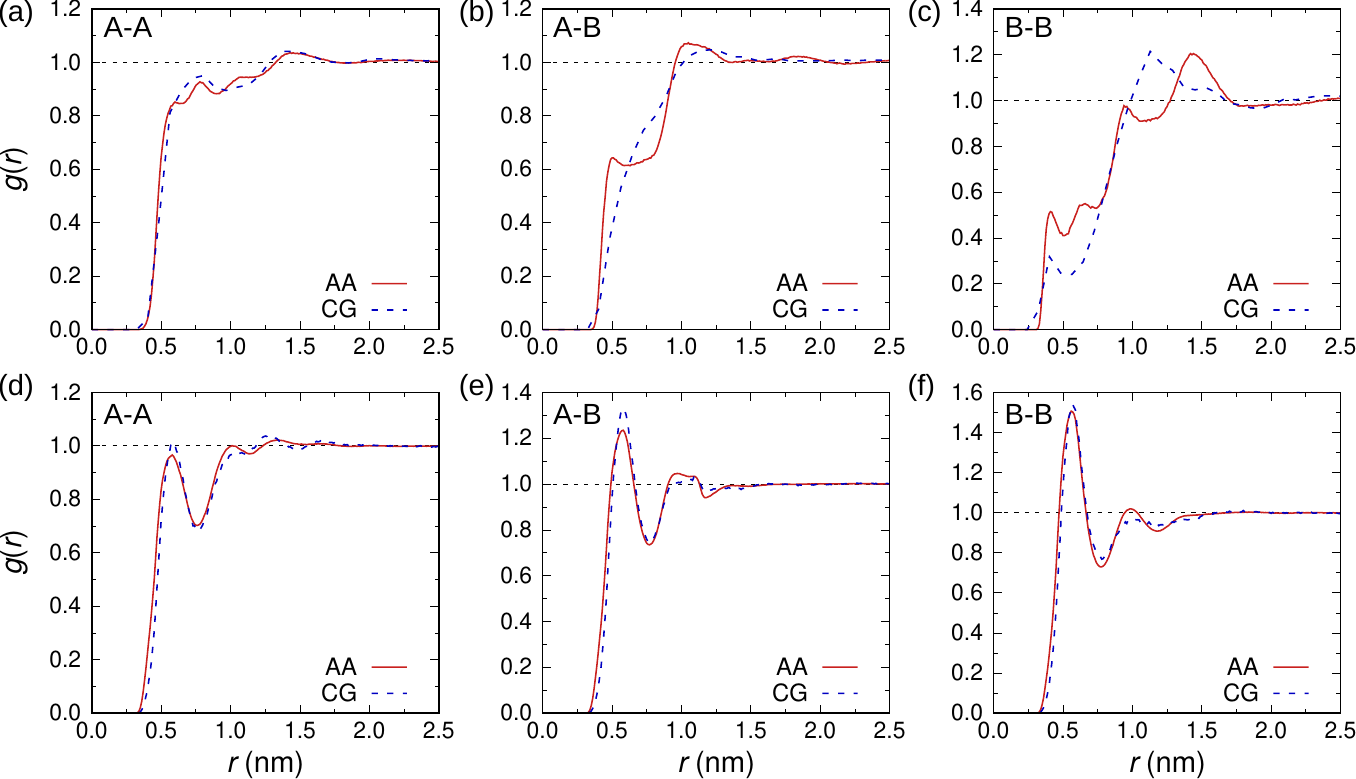}
    \caption{Comparison of radial distribution functions $g(r)$ from CG and mapped AA simulation for (a-c) TPBi and (d-e) Tm3PyPB at $T=550\,\text{K}$.}
    \label{fig:TPBi_Tm3PyPB}
\end{figure*}

Having established the CG procedure for mCBP, we next turn to other representative molecules, beginning with TPBi. In this case, we have used a low resolution scheme with two bead types [Fig.~\ref{fig:structures}(d)], and included density dependence only for the A particles (TCTA was treated in the same way, (see Sec.~S8 in SI)). This choice is again motivated by the strong spatial correlation between the central B site and its surrounding A sites. We parameterized the model at $T = 550\,\text{K}$ and determined the optimum cutoff at $r_\text{c}^* = 1.3\,\text{nm}$, which coincides with the end of second peak of the mapped A–A $g(r)$ [Figure~\ref{fig:TPBi_Tm3PyPB}(a)]. Importantly, the inclusion of this  density dependence preserves not only the A–A structure but also the A–B and B–B correlations [Fig.~\ref{fig:TPBi_Tm3PyPB}(b,c)], while also reproducing the average mass density $\bar{\rho}_\text{m}$ with high accuracy (Table~\ref{tab:bulk_density_comparison}).

For Tm3PyPB, we employed a high resolution scheme with two bead types [Fig.~\ref{fig:structures}(g)]. As with TCTA and TPBi, we first introduced a density dependence only for the A particles, but unlike those cases, this approach did not yield good agreement of the intermolecular $g(r)$ across the scanned $r_\text{c}$ range [see Fig.~S13 in the SI]. Further, the resulting bulk density $\bar{\rho}_\text{m}^\text{CG}$ determined at $r_\text{c} = 1.5\text{nm}$ was still about 44\% lower than the target density $\bar{\rho}_\text{m}^\text{AA} = 1.005\,\text{g}/\text{cm}^3$. This discrepancy suggests that the cohesive strength provided by a single A–A LD term was insufficient, because  the B sites in Tm3PyPB are located at the molecular periphery rather than buried in the core as in TCTA or TPBi. As a result, the A and B particles have weaker spatial correlations, so that the density dependence derived solely from A–A interactions cannot capture the substantial cohesive contributions involving the exposed B sites. Incorporating cross-pair density terms for A–B and B–A interactions,  significantly improved the description of all three pair structures [Fig.~\ref{fig:TPBi_Tm3PyPB}(d-f)] as well as the bulk density, $\bar{\rho}_\text{m}^\text{CG} = 0.938\,\text{g}/\text{cm}^3$. 
In Table~\ref{tab:bulk_density_comparison}, we have summarized the average mass densities $\bar{\rho}_\text{m}$ and isothermal compressibilities $\kappa$ of all investigated molecules at $T=550\,\text{K}$. The relative error $\left|\bar{\rho}_\text{m}^\text{CG}/\bar{\rho}_\text{m}^\text{AA}-1\right|$ is at most 5\% for the two-site model of mCBP, and significantly lower in most other cases. This table also summarizes the isothermal compressibility ($\kappa$) from AA and CG simulations, which we determined from constant $NPT$ simulations employing linear regression over the sampled volume and internal pressure. 

\begin{table}[ht]
    \centering
    \begin{tabular}{lcccc}
    \hline
    System & $\bar{\rho}_\text{m}^\text{AA}$ ($\text{g}/\text{cm}^3$) & $\bar{\rho}_\text{m}^\text{CG}$ ($\text{g}/\text{cm}^3$) & $\kappa^\text{AA}$ ($\times 10^{-5}$ bar$^{-1}$) & $\kappa^\text{CG}$ ($\times 10^{-5}$ bar$^{-1}$)\\
    \hline
    mCBP (2-site) & 1.042 & 0.988 & 6.05& 6.51\\
    mCBP (8-site) & 1.042 & 1.057 & 6.05& 3.4\\
    CBP & 1.022& 1.031& 10.4& 12.8\\
    BCzPh & 1.023 & 1.037& 11.2& 8.55\\
    TPBi & 1.021& 1.023& 18.0& 12.9\\
    TCTA & 1.046 & 1.043 & 9.64& 8.54\\
    Tm3PyPB & 1.005 & 0.938& 29.1 & 10.1 \\
    \hline
    \end{tabular}
    \caption{Average mass density $\bar{\rho}_\text{m}$ and isothermal compressibility $\kappa$ from AA and CG simulations at $T=550\,\text{K}$. Relative errors are below $0.1\,\%$ in all cases.}
    \label{tab:bulk_density_comparison}
\end{table}

Finally, we assessed the transferability of our CG approach by parameterizing at a single temperature $T_\text{param}$ and then performing simulations at different temperatures $T$ (Table~\ref{tab:transferability}). To this end, we considered three representative cases: the low resolution mCBP model with a single bead type, the low resolution TPBi model with two bead types and one density-dependent term, and the high resolution Tm3PyPB model with two bead types and two density-dependent terms. The agreement between the bulk number densities obtained from CG and AA simulations is generally very good, with relative errors below 2\% for mCBP and TPBi, which is comparable to the deviations already present when comparing AA and CG simulations at $T = T_\text{param}$ (see Table~\ref{tab:bulk_density_comparison}).
We have also compared the radial distribution functions from the CG and AA simulations at different temperatures $T$ (see Figs. S4, S9, and S13 in the SI).

\begin{table}[ht]
\centering
\begin{tabular}{lcccc}
\hline
System & $T_\text{param}$ (K) & $T$ (K) & $\bar{\rho}_\text{m}^\text{AA}$ ($\text{g}/\text{cm}^3$) & $\bar{\rho}_\text{m}^\text{CG}$ ($\text{g}/\text{cm}^3$) \\
\hline
\multirow{2}{*}{mCBP (2-site)}  & 680 & 650 & 0.963 & 0.983 \\
                           &  & 600 & 1.007  & 0.986 \\
\hline
\multirow{2}{*}{TPBi}  & 550 & 600 & 0.842& 0.992 \\
                           &  & 500 & 1.057 & 1.042 \\
\hline
\multirow{2}{*}{Tm3PyPB} & 550 & 500 & 1.048 & 1.023\\
                             &  & 600 & 0.957 & 0.920\\
\hline
\end{tabular}
\caption{Average mass density $\bar{\rho}_\text{m}^\text{CG}$ from CG simulations parameterized at $T_\text{param}$ and performed at $T$.}
\label{tab:transferability}
\end{table}

\subsection{Film properties}
\label{sec:results.film}
Finally, we extend our CG approach to inhomogeneous environments, focusing on free-standing liquid-vapor films. Thin films are particularly important for applications such as physical vapor deposition, where film stability, interfacial tension, and molecular orientation strongly influence device performance. In principle, CG models parameterized from bulk simulations can be directly applied to inhomogeneous systems (or vice-versa),\cite{delyser2019analysis, delyser2020interface} but here we chose to re-optimize the effective potentials using thin-film AA simulations to better capture small-$\rho$ contributions to $U_\rho$ arising near the liquid-vapor interface. To this end, we performed AA simulations at $T=550\,\text{K}$ in the $NVT$ ensemble (see Sec.~\ref{sec:methods} for details), and determined $U_2$ and $U_\rho$ by force matching, using the same optimized cutoff radii $r_\text{c}^*$ from our bulk simulations (see Sec.~\ref{sec:results.bulk}). 

\begin{figure*}[htb]
    \centering
    \includegraphics[width=\linewidth]{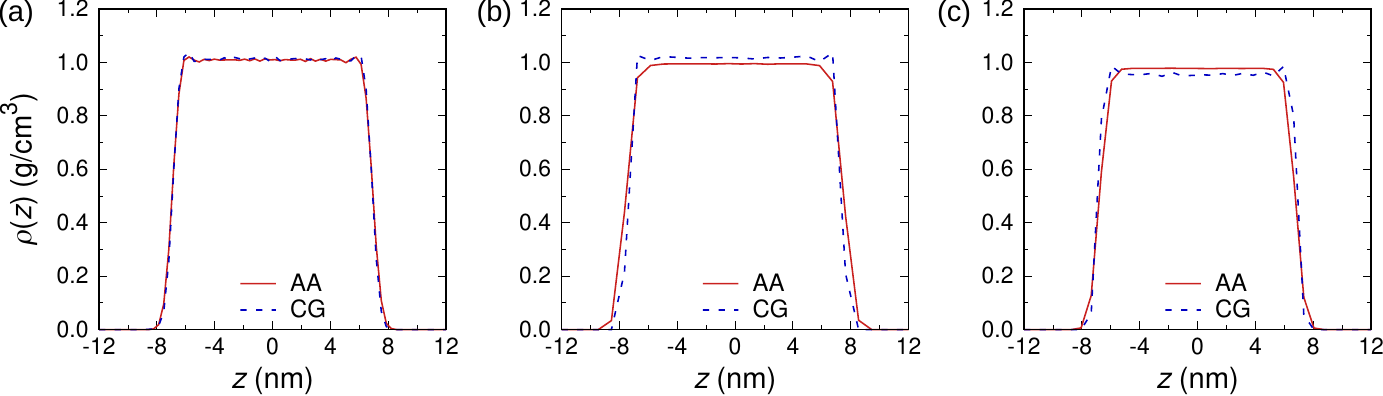}
    \caption{Density profiles $\rho_\text{m}(z)$ of (a) mCBP (two-site model), (b) TPBi, and (c) Tm3PyPB from AA and CG simulations at $T=550\,\text{K}$.}
    \label{fig:slabs}
\end{figure*}

Figure~\ref{fig:slabs} compares the density profiles $\rho_\text{m}(z)$ from AA and CG simulations for three representative molecules. The two-site CG model of mCBP reproduces $\rho_\text{m}^\text{AA}(z)$ almost perfectly [Fig.~\ref{fig:slabs}(a)], including the width of the liquid-vapor interface. The four-site TPBi model with two bead types overestimates the liquid density [Fig.~\ref{fig:slabs}(b)], likely reflecting the challenges of accurately reproducing the spatial correlations between the A-B and B-B pairs [see $g(r)$ in Fig.~\ref{fig:TPBi_Tm3PyPB}(a-c)]. In contrast, the multi-site Tm3PyPB model underestimates the liquid density only slightly [Fig.~\ref{fig:slabs}(c)], consistent with the close agreement of the bulk mass densities $\bar{\rho}_\text{m}$ (Table~\ref{tab:bulk_density_comparison}) and pair correlations $g(r)$ [Fig.~\ref{fig:TPBi_Tm3PyPB}(d-e)]. 

Many of the molecules considered in this work exhibit strong orientational preferences at liquid-vapor interfaces above $T_\text{g}$. To quantify this ordering, we introduce the second Legendre polynomial
\begin{equation}
    P_2(z) = \frac{1}{2} \left[3(\boldsymbol{\mu}(z)\cdot\mathbf{n}_z)^2 - 1\right]
\end{equation}
where $\boldsymbol{\mu}(z) \cdot \mathbf{n}_z \equiv \cos\theta$ defines the angle between the normal vector of the liquid-vapor interface, $\mathbf{n}_z$, and the characteristic director of the molecule at position $z$, $\boldsymbol{\mu}(z)$. The order parameter ranges from $P_2 = 1$ for perfect parallel alignment of $\boldsymbol{n}_z$ and $\boldsymbol{\mu}$, to $P_2 = -0.5$ if $\mathbf{n}_z$ and $\boldsymbol{\mu}$ are perpendicular. Values of $P_2 = 0$ indicate isotropic configurations without any preferred orientation.

For the AA representations of mCBP, CBP and BCzPh, $\boldsymbol{\mu}$ is the vector connecting the two nitrogen atoms. The other four molecules have a more planar geometry, and we took $\boldsymbol{\mu}$ perpendicular to the plane of the central phenyl ring (TPBi and Tm3PyPB). For TCTA, $\boldsymbol{\mu}$ corresponds to the normal of the best-fit molecular plane. We used the same conventions for the CG models: for mCBP, CBP, and BCzPh in low-resolution form, $\boldsymbol{\mu}$ is the bond vector between CG beads; for high-resolution mCBP, it is the vector between the nitrogen beads; for TPBi, Tm3PyPB, and TCTA, $\boldsymbol{\mu}$ is the normal to the plane fitted through the respective CG beads. 

\begin{figure*}[ht]
    \centering
    \includegraphics[width=1.0\linewidth]{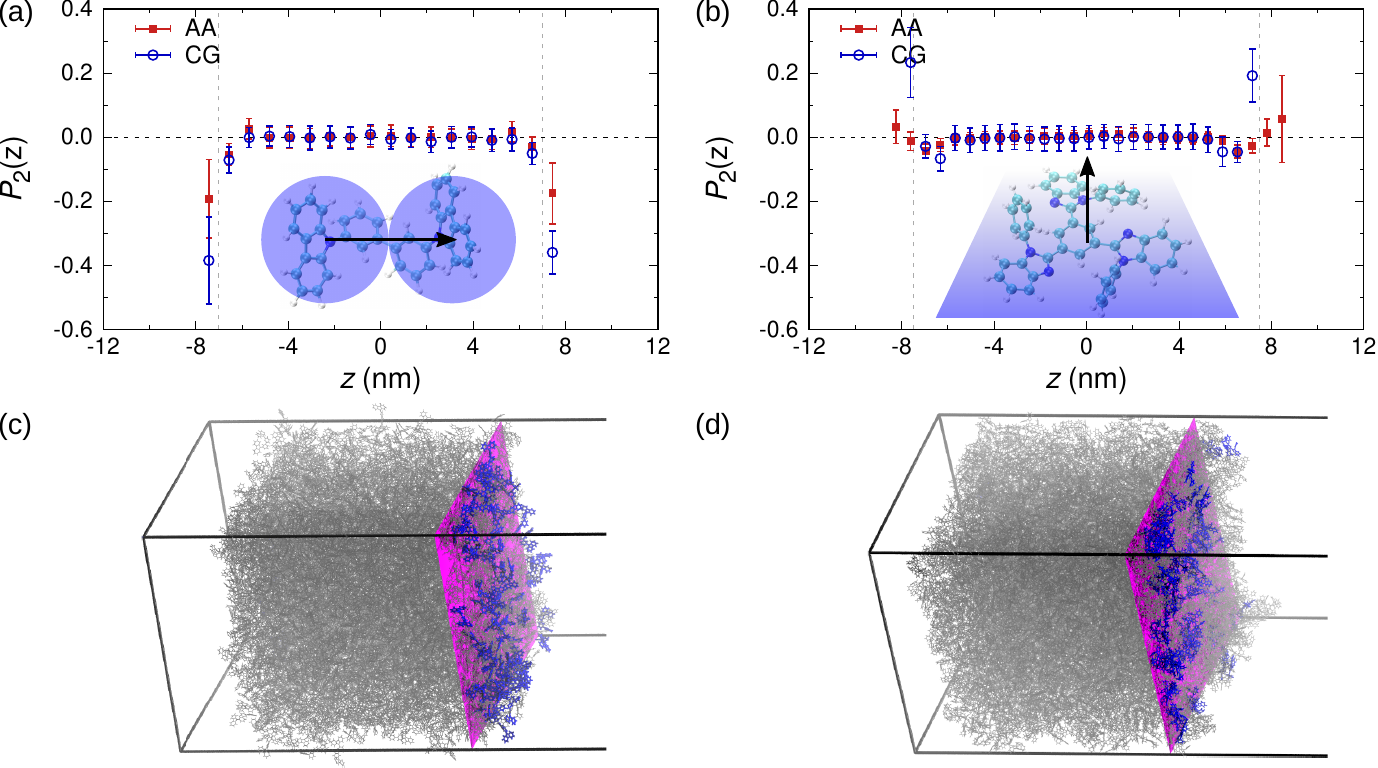}
    \caption{Local orientational order parameter $P_{2}(z)$ for (a) two-site mCBP and (b) TPBi model from AA and CG simulations vs $z$ coordinate along to interface normal. The insets indicate the molecular axis $\boldsymbol{\mu}$ that was taken for computing $\cos(\theta)$. (c) AA simulation snapshots of mCBP, where interfacial molecules with $P_2 < 0$ have been colored blue. The liquid-vapor interface is indicated by the magenta plane. (d) Same as (c) for TPBi with $P_2 > 0$ colored in blue.}
    \label{fig:orientation}
\end{figure*}

\begin{figure*}[ht]
    \centering
    \includegraphics[width=7.5cm]{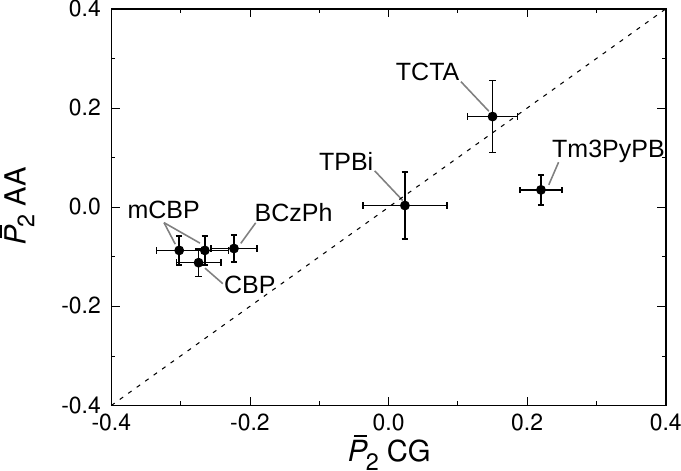}
    \caption{(a) Interfacial orientation of the molecules $\bar{P_2}$ from the CG ($x$-axis) and AA ($y$-axis) simulations. All data for $T=550\,\text{K}$.}
    \label{fig:orientation_surface}
\end{figure*}

Figure~\ref{fig:orientation} shows $P_2(z)$ profiles for mCBP and TPBi films, chosen as two representative examples of of linear and planar molecules, respectively. In the bulk-like interior of the films, both systems are isotropic with $P_2 \approx 0$. At the vapor-liquid interface, however, their behavior diverges. For mCBP, $P_2$ becomes strongly negative, indicating a lying down orientation of the molecules. Here, the ordering is slightly more pronounced in the AA simulations compared to the CG simulations, but both capture the same trend. In contrast, the TPBi molecules exhibit slightly positive $P_2$ values at the interface, reflecting a tendency for the molecular planes to align parallel to the surface. Figure~\ref{fig:orientation_surface} shows a comparison of the interfacial $\bar{P_2}$, computed by averaging $P_2(z)$ over the liquid-vapor interface (see Sec.~S11 in SI), highlighting the close agreement between the AA and CG simulations for the majority of molecules.

We further characterized the interfacial properties of the thin films through their surface tension $\gamma$, which we computed using the Kirkwood-Buff relation\cite{kirkwood1949statistical}
\begin{equation}
\gamma = \frac{L_z}{2} \left\langle P_{zz} - \frac{1}{2}(P_{xx} + P_{yy}) \right\rangle
\end{equation}
where $P_{ii}$ are the diagonal components of the pressure tensor along the Cartesian $i = x, y, z$ directions. Table~\ref{tab:gamma_comparison} compares the surface tensions obtained from the AA and CG slab simulations. All CG models systematically overestimate $\gamma$ relative to their AA counterparts by a factor of 2 to 4. This trend is somewhat counterintuitive in light of previous works on density-independent CG models, where smoother effective interactions often lead to reduced surface tensions at liquid–vapor interfaces.\cite{ndao2015gamma, li2017FNP, bauer2022IDP} One possible explanation for these systematic differences could be how the LDPs alter the distribution of cohesive interactions between the bulk and interface regions; although our CG models reproduce bulk coexistence densities and compressibilities, they may effectively strengthen cohesion in dense regions and penalize low-density regions more than the underlying AA models. As a result, the free-energy cost of sampling sub-bulk densities at the liquid–vapor interface could be increased, leading to higher $\gamma$ values.

\begin{table}[ht]
    \centering
    \begin{tabular}{lcc}
    \hline
    System & $\gamma^\text{AA}$ ($\text{mJ}/\text{m}^2$) & $\gamma^\text{CG}$ ($\text{mJ}/\text{m}^2$)\\
    \hline
    mCBP (2-site) & $18 \pm 4$ & $50.7 \pm 0.2$\\
    mCBP (8-site) & $18 \pm 4$ & $77.5 \pm 3.5$\\
    CBP & $21 \pm 4$ & $47.1 \pm 0.1$\\
    BCzPh & $10 \pm 6$ & $36.8 \pm 0.1$\\
    TPBi & $17 \pm 3$ & $33.4 \pm 0.1$\\
    TCTA & $32 \pm 8$ & $53.2 \pm 0.5$\\
    Tm3PyPB & $17 \pm 3$ & $44.5 \pm 0.7$\\
    \hline
    \end{tabular}
    \caption{Interfacial tension $\gamma$ from AA and CG simulations at $T=550\,\text{K}$.}
    \label{tab:gamma_comparison}
\end{table}

\subsection{Computational efficiency}
\label{sec:results.speedup}

Finally, we assessed the computational efficiency of our CG approach by comparing AA and CG simulations of bulk mCBP under identical $NVT$ conditions. In general, the LDP can be evaluated as an effective pair interaction, so that it can benefit from the typical optimization strategies as cell and Verlet neighbor lists. If the cutoff radius, $r_\text{c}$, of the weight function, $w(r)$, for a given pair of particle types is smaller than or equal to that of the corresponding pair potential $U_2$, the same neighbor list can be reused for both contributions. This reuse significantly amortizes the additional cost associated with computing local densities, such that the LDP adds only a modest overhead relative to standard CG pair potentials.

Coarse-graining affords two distinct sources of speed-up. First, the number of time steps per wall-clock second (TPS) can increase because the number of interaction sites (and corresponding pairwise evaluations) is reduced. Second, the effective dynamics become faster in physical time, so that each integration time step corresponds to a longer segment of real time. We quantify both effects below through simulations of the AA model and the two- and eight-site CG models of mCBP. All simulations were performed using the same computational resources (Intel Xeon Platinum 8260 CPUs), with parallelization achieved via domain decomposition using MPI. A caveat of our comparison is that different simulation packages were employed for the AA and CG simulations (GROMACS for AA, LAMMPS for CG), so that some differences in raw performance may reflect code-level rather than model-level effects. The number of molecules was fixed to 3,000 in all cases, corresponding to 6,000 particles for the two-site CG model, 24,000 particles for the eight-site CG model, and 186,000 particles for the AA model.

For the two-site CG model, we achieved about $16\,\text{TPS}$ on a single CPU core, which increased to $260\,\text{TPS}$ when using 20 CPU cores. Here, the parallel efficiency $E$ -- defined as the ratio between the observed speedup and the number of CPU cores -- decreased with increasing number of CPU cores from $E \approx 0.91$ for 4 cores to $E \approx 0.80$ for 20 cores. This decrease in efficiency is likely due to the increasing communication overhead and under-utilization of each core that occurs with decreasing domain size. The eight-site model is significantly slower, achieving about $1.8\,\text{TPS}$ on a single CPU core and about $27\,\text{TPS}$ on 20 cores ($E \approx 0.75$). For comparison, the AA simulations achieve about $57\,\text{TPS}$ on 20 cores, which lies in between the performance of the two- and eight-site CG models.

To take into account the accelerated dynamics in physical time, we first computed the self-diffusion coefficient from the mean-square displacement $\la \Delta r^2 \ra$ of the molecules in the three representations, and then determined the average number of time steps needed for an mCBP molecule to diffuse over $\la \Delta r^2 \ra = 1\,\text{nm}^2$. The AA simulations required about $10^6$ time steps, while the eight-site and two-site model require about $9\times10^4$ and $10^4$ steps, respectively. Taking the TPS values for simulations running on 20 cores, we find that the AA simulations need almost 5 hours. In contrast, the eight-site CG models requires approximately 50 minutes to reach the same physical time, while the two-site CG model needs only about 40 seconds.

\section{Conclusions}
We developed coarse-grained models for small organic molecules, which accurately capture both bulk and interfacial behavior through local-density-dependent potentials (LDPs). The coarse-grained models accurately reproduce key bulk properties, including density, compressibility, and radial pair distribution functions, and exhibit satisfactory transferability over a broad range of temperatures. Further, they capture strong density fluctuations within a single system, which is essential for describing inhomogeneous situations such as thin films and droplets. Finally, the combination of a reduced number of interaction sites with smoother effective energy landscapes yields an overall computational speed-up of approximately two orders of magnitude relative to fully atomistic simulations.

These results suggest several promising directions for future work. Transferability could be further improved by interpolating potentials parametrized at different temperatures, enabling a more systematic description of thermodynamic state dependence. In addition, the demonstrated ability to resolve interfacial structure and large density variations at greatly reduced cost makes this approach particularly well suited for studying vapor deposition and related non-equilibrium growth processes, which remain challenging to access with fully atomistic simulations at relevant time and length scales.

\appendix
\section{Atomistic Simulation Details}
\label{sec:methods}
All Lennard–Jones parameters, as well as bonded parameters of AA force-field were taken from OPLS-AA with a cutoff radius of \SI{13}{\angstrom} for the non-bonded interactions.\cite{jorgensen2005opls, jorgensen1996development} Electrostatic interactions were treated with the Particle Mesh Ewald method using a Fourier spacing of \SI{0.12}{\nano\meter}. Bulk simulations were carried out in the $NPT$ ensemble with 3000 molecules packed in a periodic cubic box, while liquid-vapor slab simulations were performed at constant $NVT$. The temperature was maintained with a velocity-rescaling thermostat\cite{bussi2007canonical} (\SI{0.5}{\pico\second} time constant) and the pressure at $P = 1\,\text{bar}$ with a Berendsen barostat\cite{berendsen1984molecular} (\SI{0.5}{\pico\second} time constant, compressibility \SI{4.5e-5}{\per\bar}). The timestep was \SI{1}{\femto\second}, and each AA simulation was \SI{20}{\nano\second} long, with samples taken every \SI{1}{\pico\second}. All simulations were conducted above the glass transition temperature $T_\text{g}$, which we determined from bilinear fits to density–temperature curves (Table~\ref{tab:Tg}).\cite{lin2021glass} 

\begin{table}[htbp]
\centering
\begin{tabular}{ccc}
\hline
System &  $T_\text{g}^\text{AA}$ (K) & $T_\text{g}^\text{exp}$ (K)\\
\hline
mCBP & $422 \pm 5$& 366\cite{mCBP_exp} \\
CBP & $393 \pm 9$&  388\cite{CBP_exp}\\
BCzPh & $386 \pm 7$& 371\cite{BCzPh_exp}\\
TPBi & $472 \pm 9$& ---\\
TCTA & $429 \pm 2$& 427\cite{TCTA_exp}\\
Tm3PyPB & $377 \pm 9$& ---\\
\hline
\end{tabular}
\caption{Comparison of $T_\text{g}$ from AA simulations and experiments. Error bars for $T_\text{g}^\text{AA}$ reflect uncertainty from bilinear fits to density-temperature curves.\cite{lin2021glass}}
\label{tab:Tg}
\end{table}

\section*{Data availability}
The data that support the findings of this study are available from the corresponding authors upon reasonable request.

\section*{Conflicts of interest}
There are no conflicts to declare.

\section*{Acknowledgments}
This work was supported by the Deutsche Forschungsgemeinschaft (DFG, German Research Foundation) through the framework of the collaborative research center TRR 146 (Project No. 233630050), and through the Heisenberg grant to A.N. (Project No. 470113688). The authors acknowledge the Center for High-Performance Computing (ZIH) Dresden for providing computational resources. M.C.L. and W.G.N. acknowledge financial support from the US National Science Foundation (Grant Nos. CHE-2154433 and CHE-2502094).

\providecommand{\latin}[1]{#1}
\makeatletter
\providecommand{\doi}
  {\begingroup\let\do\@makeother\dospecials
  \catcode`\{=1 \catcode`\}=2 \doi@aux}
\providecommand{\doi@aux}[1]{\endgroup\texttt{#1}}
\makeatother
\providecommand*\mcitethebibliography{\thebibliography}
\csname @ifundefined\endcsname{endmcitethebibliography}
  {\let\endmcitethebibliography\endthebibliography}{}

\end{document}


\maketitle
\clearpage

\section{Details of the AA force-fields}
Parameters for the non-bonded interactions were taken from Ref.~\citenum{lin2021glass}, following the method of Cole {\it et al.}\cite{cole2016biomolecular} There, overlapping atomic electronic densities were obtained via the DDEC6 scheme;\cite{manz2016introducing} atomic partial charges were computed using the Merz–Kollmann method by fitting the electrostatic potential from density functional theory calculations at the B3LYP/6-31G(d) level.\cite{singh1984approach} Dihedral potentials were parametrized by fragmenting molecules.\cite{mondal2021molecular} 

All simulations contained $3,000$ molecules, and were initialized by randomly inserting the molecules without overlap into a cubic box at low density (\SIrange{450}{500}{\kilo\gram\per\meter\cubed}). Then, the systems were heated from \SI{100}{\kelvin} to \SI{300}{\kelvin} at \SI{0.67}{\kelvin\per\pico\second}, followed by a \SI{2}{\nano\second} equilibration. The system was then further heated to \SI{800}{\kelvin} at \SI{0.5}{\kelvin\per\pico\second} and equilibrated for \SI{10}{\nano\second}, then cooled to \SI{0}{\kelvin} at \SI{100}{\kelvin\per\nano\second}. 

\section{Bonded interactions of the CG force fields}
Based on different CG mapping schemes, we have defined different types of bonded interactions for the different molecules considered in this study (see Table~\ref{tab:bonded}). We parameterized these potentials via Boltzmann inversion (BI) using the VOTCA (``Versatile Object-Oriented Toolkit for Coarse Graining'')  software package.\cite{ruhle2009versatile} In BI, the effective potential is computed by inverting the probability distribution $P$
\begin{equation}
    U(q) = -k_\text{B} T \ln \left[ P(q) \right]
\end{equation}
where $q = r$, $\theta$ or $\phi$, depending on the bonded interaction. Correlations between different degrees of freedom were neglected, so that we treated the bond, angle, and dihedral distributions independently. The resulting effective interaction potentials were then smoothed and tabulated for use in the CG simulations.

\begin{table}[htbp]
\centering
\begin{tabular}{cc}
\hline
\textbf{Molecule} & \textbf{CG bonded interactions} \\
\hline
mCBP, CBP, BCzPh (2-site) & Bond A–A \\
TPBi, TCTA & Bond A–B; Angle A–B–A; Dihedral B–A–A–A \\
Tm3PyPB & Bond A–B, A–A; Angle A–B–A, A–A–A;\\
                               & Dihedral B–A–A–A, A–A–A–A \\
mCBP (8-site)     & Bond A–A, A–B; Angle A–A–B, A–B–A;\\
                                 & Dihedral B–A–A–A, A–A–A–A \\
\hline
\end{tabular}
\caption{Bonded interactions based on mapping schemes and molecules.}
\label{tab:bonded}
\end{table}

\section{Failure of pair potential: Two-site mCBP example}
\label{sec:U2-paironly}
\begin{figure*}[ht]
\centering
\includegraphics[width=1.0\linewidth]{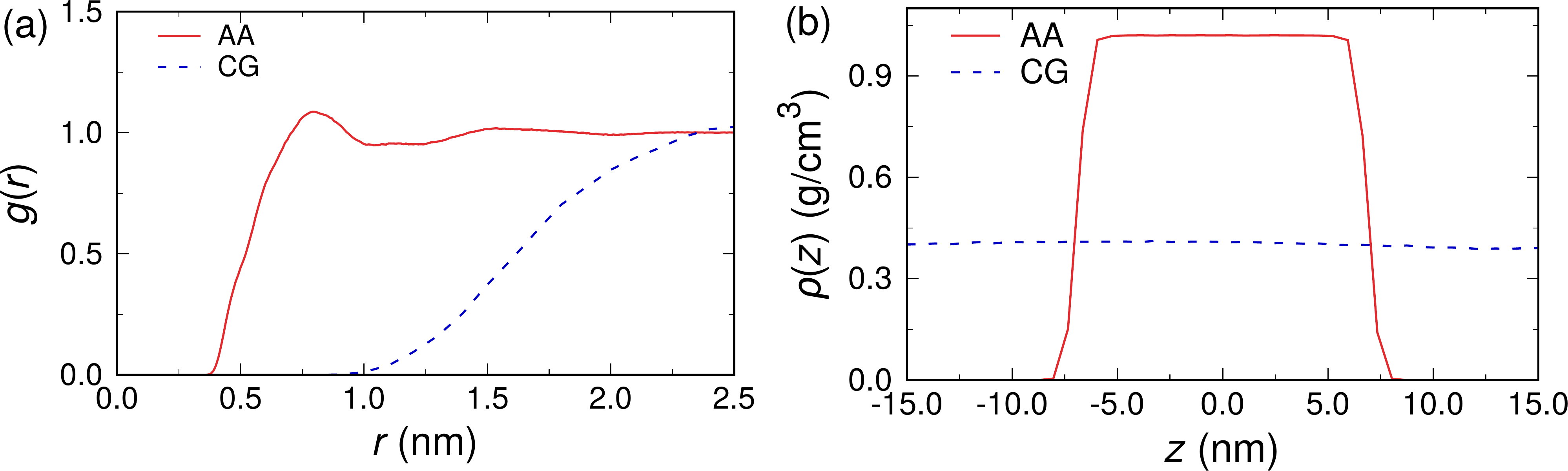}
\caption{(a) Radial distribution function, $g(r)$, from AA and CG (two-site model) simulations of mCBP. CG simulations have been performed with $U_2^\text{pair-only}$, parameterized from bulk $NPT$ simulations. (b) Corresponding density profiles $\rho(z)$ for a thin film from AA and CG simulations. All data shown for $T=550\,\text{K}$.}
\label{fig:mCBP-U2}
\end{figure*}

To better understand the effect of the local-density-dependent potential (LDP), we also parameterized the CG model using only pair-wise potentials, obtained by applying the force-matching (FM) variational principle. For the two-site mCBP CG model, the resulting interaction $U_{2}^{\text{pair-only}}$, parameterized from AA bulk simulations under $NPT$ conditions is very repulsive (see main text) and insufficient to simultaneously reproduce the bulk density and the pair-structure (Fig.~\ref{fig:mCBP-U2})  from the CG simulations under $NPT$ conditions. At constant pressure, the CG model samples a low-density vapor state.  Furthermore, the CG model is unable to maintain a stable film in $NVT$ conditions. Indeed, parameterizing $U_{2}^{\text{pair-only}}$ from AA slab simulations under constant $NVT$ condition is also insufficient for reproducing the liquid-vapor coexistence from CG simulations (Fig.~\ref{fig:slabs-U2}(a)). 

\section{Additional details for two-site mCBP model}
\subsection*{Comparison to $U_2^\text{pair-only}$}
\begin{figure}[ht]
\centering
\includegraphics[width=1.0\linewidth]{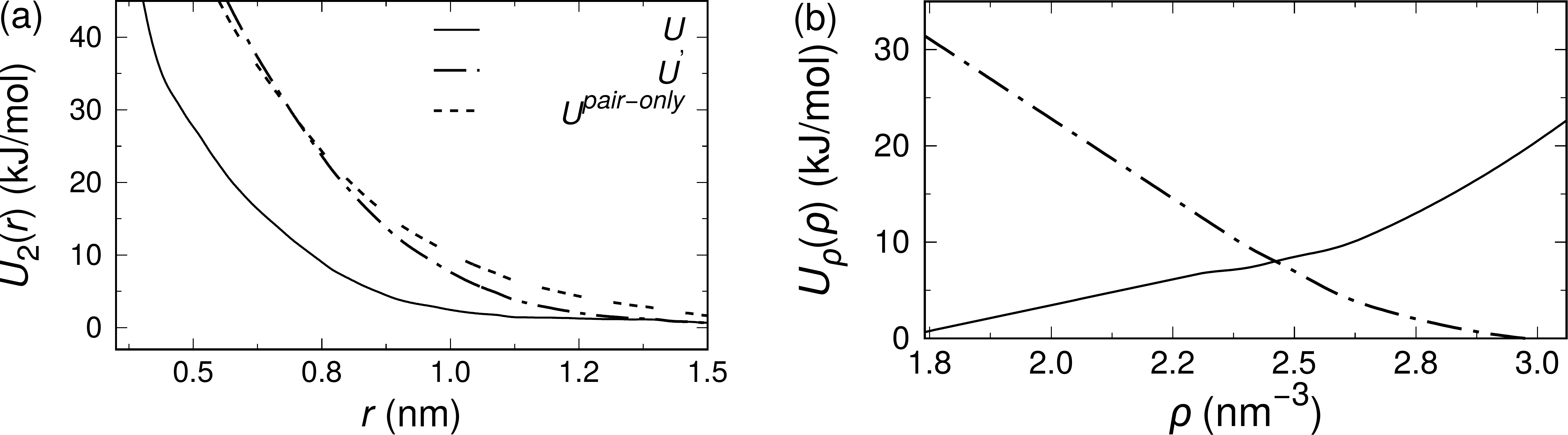}
\caption{(a) Two parameterizations of $U$, where $U_{2}^{'}$ has been been shifted to match $U_2^\text{pair-only}$ (see Sec.~\ref{sec:U2-paironly}). (b) Corresponding LDPs $U_\rho$ and $U_\rho'$. All data taken at $T=550\,\text{K}$.} 
\label{fig:U-prime}
\end{figure}

Because the total force is invariant under the simultaneous transformations $U_{2}'(r) = U_{2}(r) + c\,\bar{\omega}(r)$ and $U_{\rho}'(\rho) = U_{\rho}(\rho) - c\rho/2$, the set of potentials $U = \{U_{2}, U_{\rho}\}$ are not unique but form a one-parameter family of equivalent representations, $U^{'}(c) = \{U_{2}^{'}(c), U_{\rho}^{'}(c)\}$,  that produce identical forces and distributions.\cite{delyser2019analysis} This transformation shifts the LDP and slightly modifies the pair potential only for $r \le r_\text{c}$, where $w'(r) \neq 0$. Figure~\ref{fig:U-prime} shows two parameterizations of $U$.

\subsection*{Analysis of optimal $r_\text{c}$}
\begin{figure}[ht]
\centering
\includegraphics[width=0.9\linewidth]{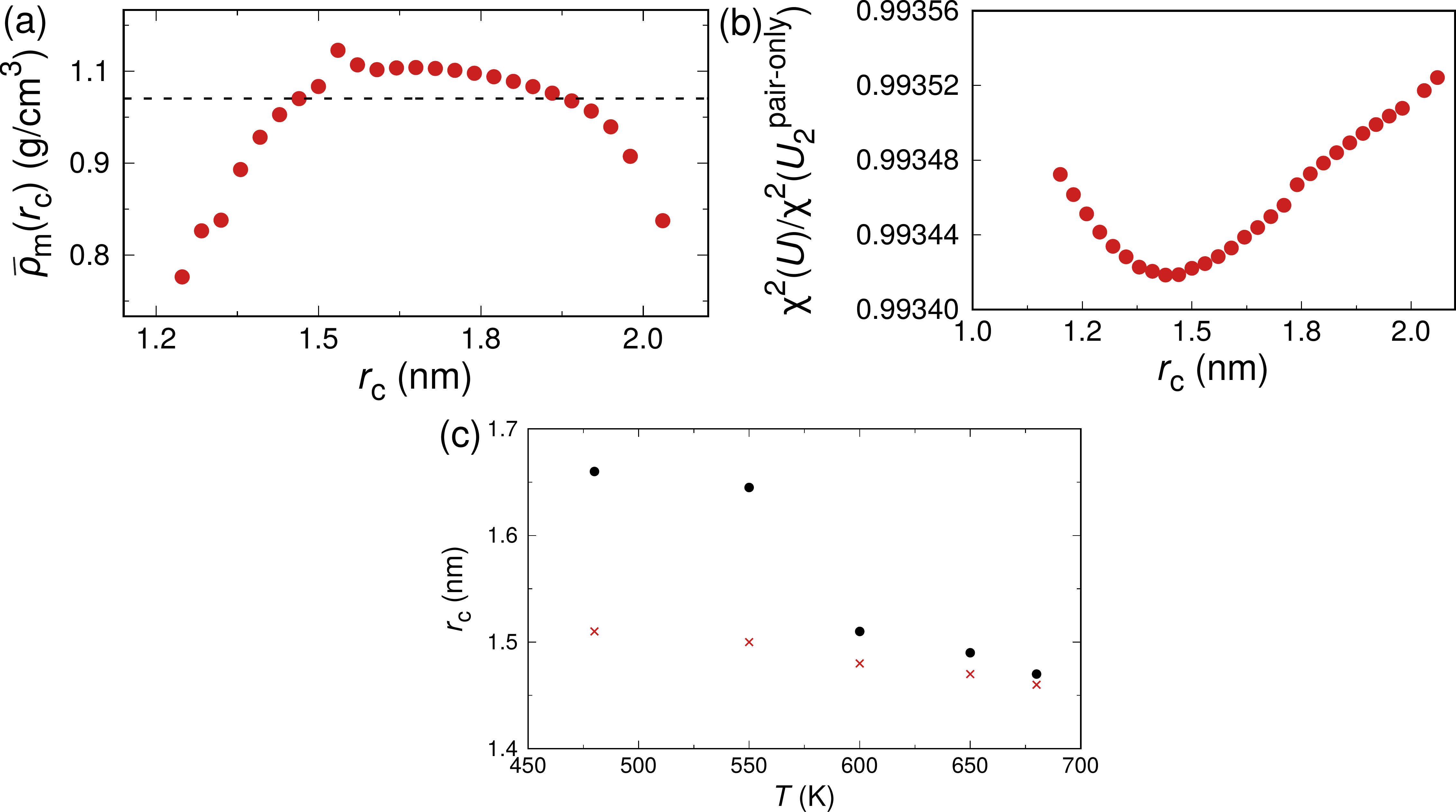}
\caption{(a) Bulk density $\bar{\rho}_\text{m}$ vs. cutoff radius $r_\text{c}$. The black dotted line represents $\bar{\rho}_\text{m}^\text{AA}$. (b) Normalized FM error $\chi^2(U)/\chi^2(U_2^\text{pair-only})$ vs. $r_\text{c}$. All data computed at $T=650\,\text{K}$. (c) Optimum $r_\text{c}^*$ (black dots) and $r_\text{c}$  where $\chi^2$ is minimized (red crosses) as a function of $T$.}
\label{fig:mCBP-cutoff}
\end{figure}

In the main text, we have determined the $T$-dependent optimal cutoff-radius of the weight function $w(r)$, $r_\text{c}^*$, through the condition $\bar{\rho}_\text{m}^\text{CG} \approx \bar\rho_\text{m}^{\text{AA}}$. Interestingly, for $T = 480\,\text{K}$ and $T = 550\,\text{K}$, $r_\text{c}^*$ is much longer than for the remaining higher temperatures (Fig.~\ref{fig:mCBP-cutoff}(c)). Figure~\ref{fig:mCBP-cutoff} shows the influence of $r_\text{c}$ on $\bar{\rho}_\text{m}^\text{CG}$ for $T = 650\,\text{K}$. Figure~\ref{fig:mCBP-cutoff}(b) shows the normalized FM error $\chi^2(U)/\chi^2(U_2^\text{pair-only})$ for the investigated range of $r_\text{c}$. 

\subsection*{Analysis of transferability of LDPs}
\begin{figure}[ht]
\centering
\includegraphics[width=0.6\linewidth]{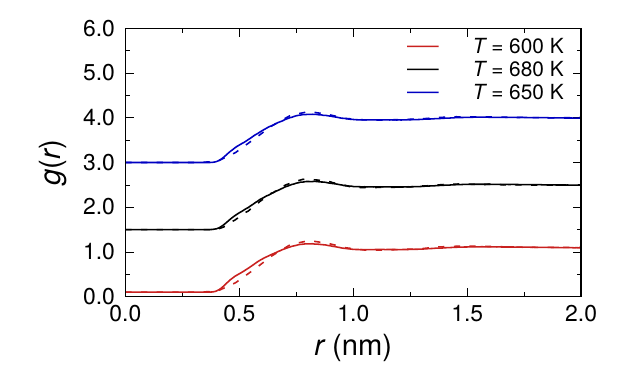}
\caption{Comparison of $g(r)$ from AA (solid lines) and CG simulations (dashed lines). CG potentials parameterized at $T=680\,\text{K}$. Curves shifted vertically for better visibility.}
\label{fig:mCBP-transferability}
\end{figure}

To further test the transferability of $U = \{U_2, U_\rho\}$, we parameterized  $U_2$ and $U_{\rho}$ at $T=680\,\text{K}$, and used these potentials for simulations at $T = 600\,\text{K}$ and $T = 650\,\text{K}$ under constant $NPT$ conditions. Figure~\ref{fig:mCBP-transferability} shows the comparison of $g(r)$ from the corresponding AA and CG simulations.

\section{Additional details for eight-site mCBP model}
\begin{figure*}[htb]
\centering
\includegraphics[width=1.0\linewidth]{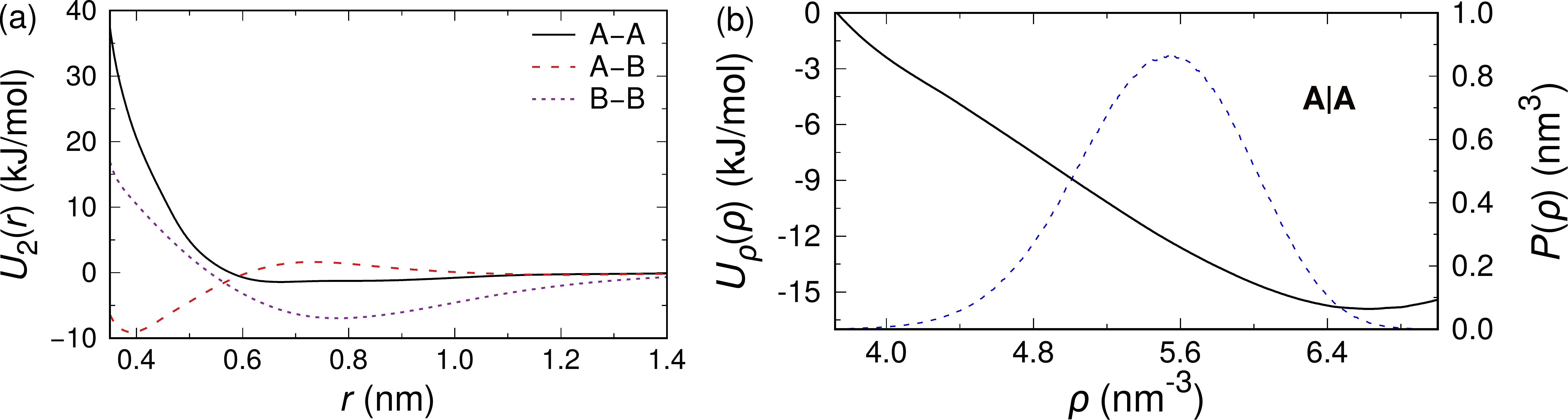}
\caption{(a) Pair potentials $U_2$ for the eight-site CG model of mCBP. Interaction for A-A pairs transformed to match $U_2^\text{pair-only}$ as closely as possible. (b) Corresponding LDP ($U_\rho$, left $y$-axis) and local-density distribution ($P(\rho)$, right $y$-axis). All data taken at $550\,\text{K}$.}
\label{fig:mCBP-8site}
\end{figure*}

The eight-site CG model of mCBP has two bead types, A and B, where the LDP considers only the local density of A particles around A sites (see main text for details). Figure~\ref{fig:mCBP-8site}(a) shows the pair potentials $U_2$, parameterized from bulk $NPT$ simulations at $T=550\,\text{K}$ and $P=1\,\text{bar}$. Further, the A-A pair potential has been transformed to match $U_2^\text{pair-only}$ as closely as possible. The corresponding LDP, $U_\rho$, and local-density distribution, $P(\rho)$ at the optimized cutoff radius $r_\text{c}^*$, are shown in Figs.~\ref{fig:mCBP-8site}(b) and \ref{fig:mCBP-8site}(c), respectively.

Table~\ref{tab:mCBP-8site} summarizes the effect of $r_\text{c}$ on the average mass density, $\bar\rho_\text{m}^\text{CG}$, and isothermal compressibility, $\kappa^\text{CG}$. 

\begin{table}[htb]
    \centering
    \begin{tabular}{lcccc}
    \hline
    $r_\text{c}$ (nm) & $\bar{\rho}_\text{m}^{\text{CG}}$ (g/cm$^3$) & $\kappa^{\mathrm{CG}}$ ($\times 10^{-5}$ bar$^{-1}$)\\
    \hline
     0.900& 0.514 & 15.2 \\
     0.950& 0.674 & 28.0 \\
     1.000& 0.819  & 10.9  \\
     1.150& 0.902 & 7.8 \\
     1.200& 0.964  & 4.7 \\
     1.250& 1.014 & 4.2 \\
     1.300& 1.057&  3.4\\
    \hline
    \end{tabular}
    \caption{Average mass density $\bar\rho_\text{m}^\text{CG}$ and isothermal compressibility $\kappa^\text{CG}$ from CG simulations at $T=550\,\text{K}$ for the eight-site CG model of mCBP. From AA simulations, $\bar{\rho}_\mathrm{m}^{\mathrm{AA}} =1.0417\,\text{g}/\text{cm}^3$ and $\kappa^\text{AA}= 6.053\times10^5\,\text{bar}^{-1}$.}
    \label{tab:mCBP-8site}
\end{table}

\begin{figure*}[ht]
\centering
\includegraphics[width=\linewidth]{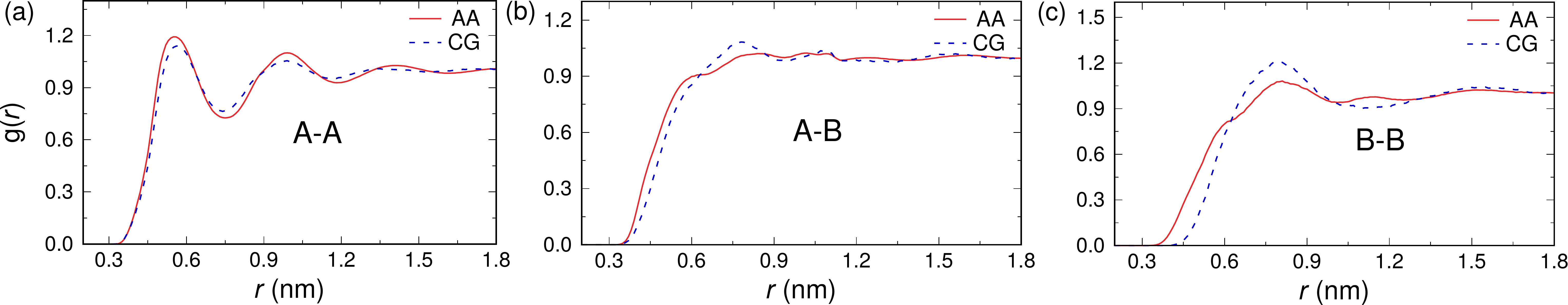}
\caption{Radial distribution functions, $g(r)$, for the (a) A-A, (b) A-B, and (c) B-B pairs of the eight-site CG model of mCBP. All simulations have been performed at $T= 550\,\text{K}$.}
\label{fig:mCBP-8-site-gr}
\end{figure*}

Finally, we have compared the radial distribution functions, $g(r)$, for the three different pair types (Fig.~\ref{fig:mCBP-8-site-gr}), revealing overall good agreement with the reference AA simulations. 

\section{Additional details for BCzPh and CBP models}
\begin{figure}[ht]
\centering
\includegraphics[width=1.0\linewidth]{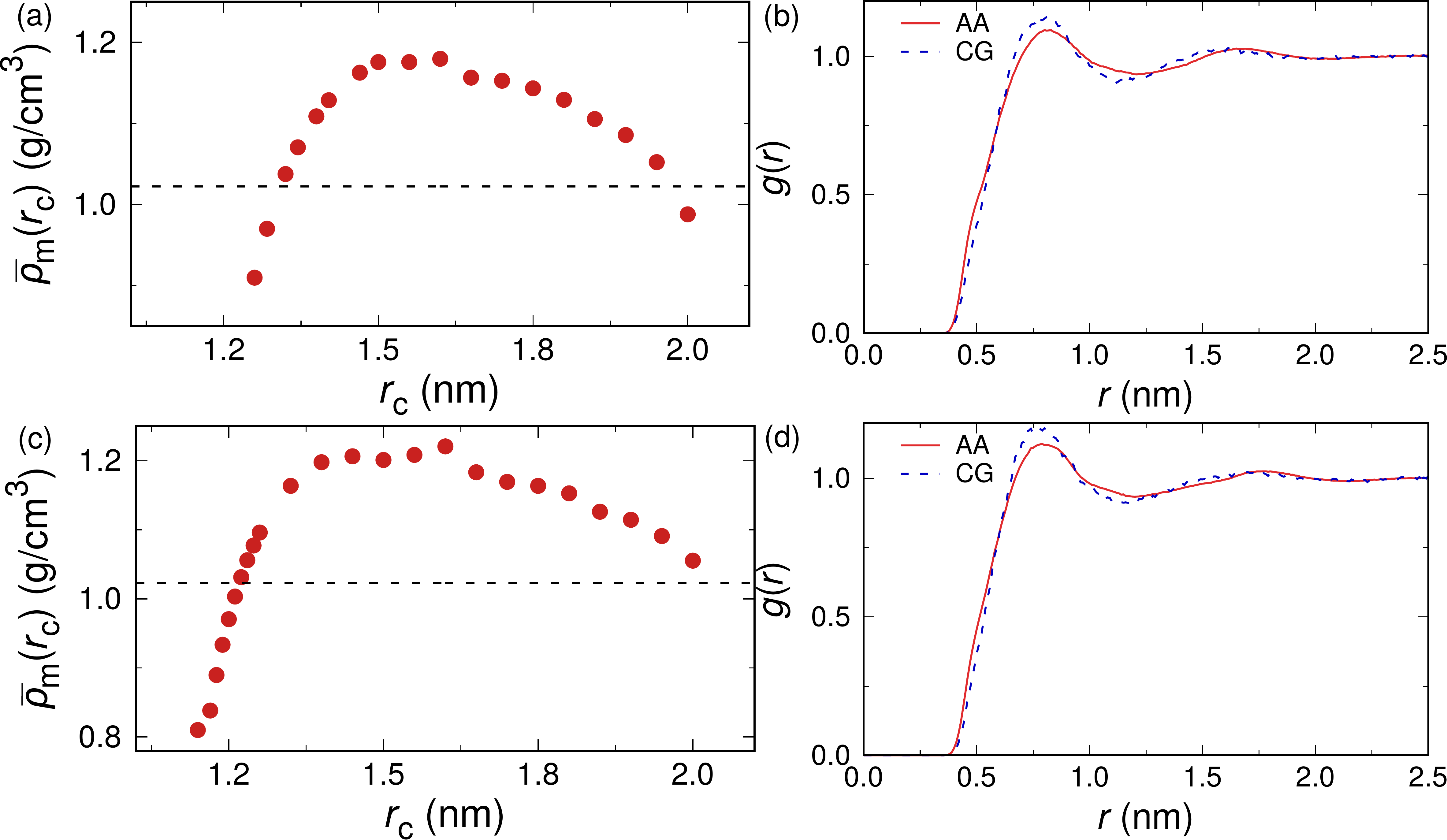}
\caption{(a) Average mass density $\bar{\rho}_\text{m}^\text{CG}$ as a function of cutoff radius $r_\text{c}$ from CG simulations of BCzPh. The dashed line indicates $\bar{\rho}_\text{m}^\text{AA}$. (b) Radial distribution function for A-A pairs from AA and CG simulations, as indicated. (c, d) Same data as in (a, b) but for CBP. All data taken at $T=550\,\text{K}$}
\label{fig:BCzPh-CBP}
\end{figure}

BCzPh and CBP were coarse-grained into two sites, like mCBP (see main text). These three molecules are fragment wise equivalent, resulting in a similar treatment of the LDPs. Figures~\ref{fig:BCzPh-CBP}(a,c) show the average mass densities, $\bar{\rho}_\text{m}^\text{CG}$ for different $r_\text{c}$ at $T= 550\,\text{K}$ for these two molecules. Figures~\ref{fig:BCzPh-CBP}(b,d) show the resulting radial distribution functions $g(r)$, from both the AA simulations and CG simulations with the optimized cutoff radius $r_\text{c}^*$, for the A-A pairs of BCzPh and CBP, respectively.

\section{Additional details for TPBi model}
\begin{figure*}[ht]
\centering
\includegraphics[width=1.0\linewidth]{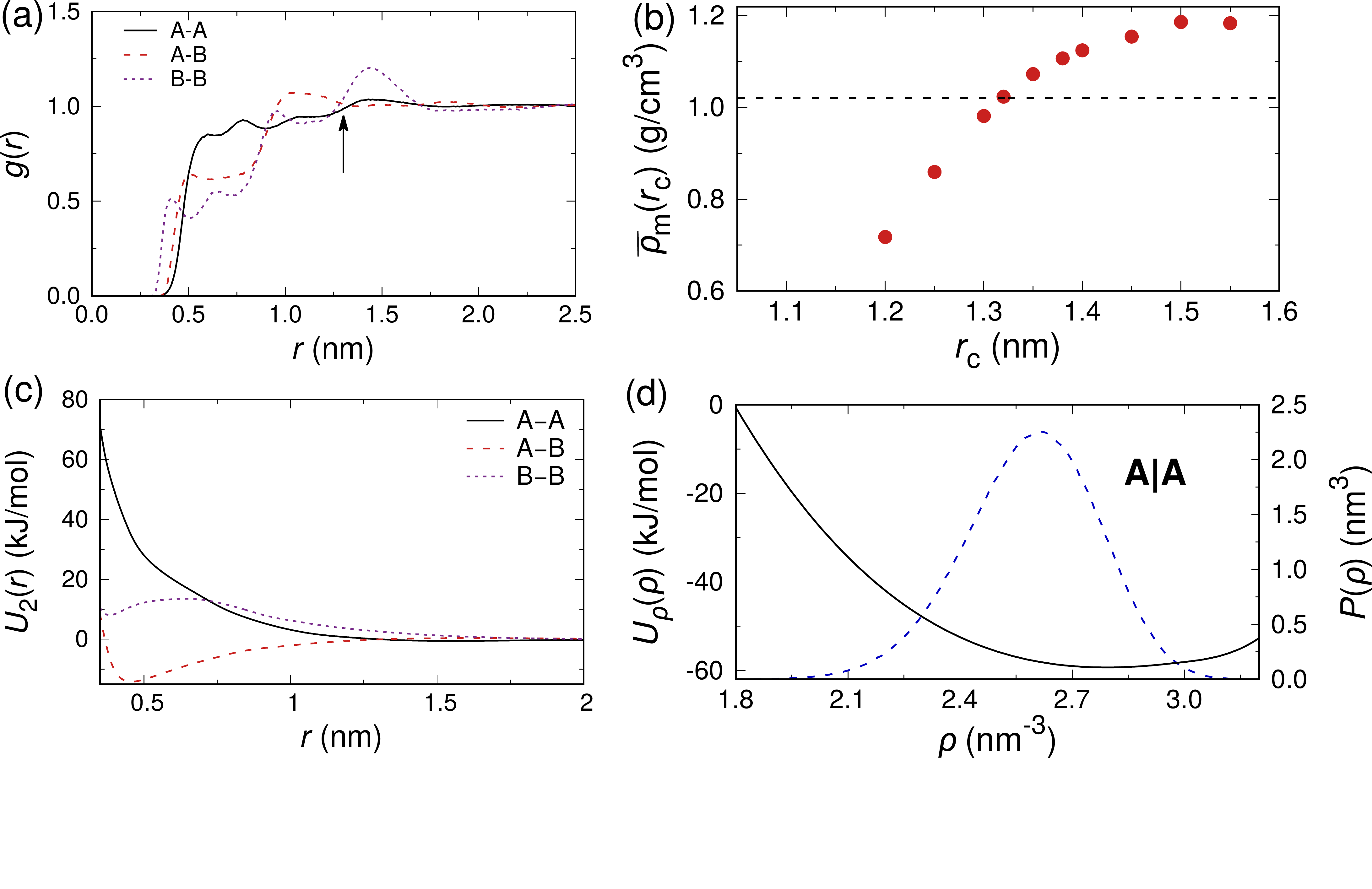}
\caption{(a) Radial distribution function, $g(r)$, from the mapped AA simulations of TPBi. (b)  Average mass density $\bar{\rho}_\text{m}^\text{CG}$ as a function of cutoff radius $r_\text{c}$ from CG simulations of TBPi. The dashed line indicates $\bar{\rho}_\text{m}^\text{AA}$. (c) Pair potentials $U_2$, where interaction for A-A pairs transformed to match $U_2^\text{pair-only}$ as closely as possible. (d) Corresponding LDP ($U_\rho$, left $y$-axis) and local-density distribution ($P(\rho)$, right $y$-axis). All data taken at $T = 550\,\text{K}$}
\label{fig:TPBi}
\end{figure*}

We mapped TBPi molecules to four-sites using two particle types, where the LD potential was only used for the A$|$A pairs. Figure~\ref{fig:TPBi}(a) shows the radial distribution functions of the A-A, A-B and B-B pairs from the mapped AA simulations. Figure~\ref{fig:TPBi}(b) shows the average mass density as a function cutoff radius, while Figs.~\ref{fig:TPBi}(c,d) show the resulting potentials $U_2$ and $U_\rho$ determined at the optimum cutoff radius $r_\text{c}^* = 1.32\,\text{nm}$. 

\begin{figure*}[ht]
\centering
\includegraphics[width=1.0\linewidth]{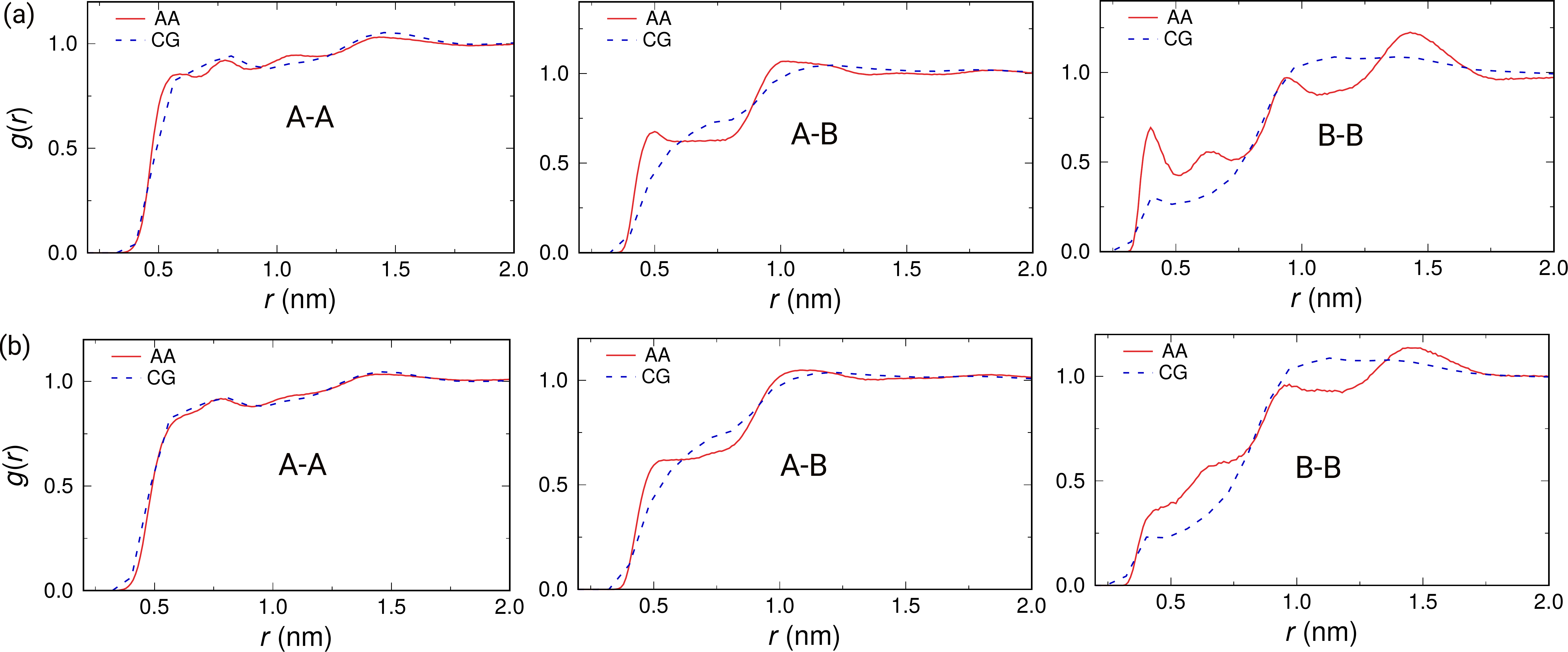}
\caption{(a) Radial distribution functions $g(r)$ for  A-A, A-B, and B-B pairs from AA and CG simulations, as indicated, at $T=500\,\text{K}$. (b) Same as (a) but for $T=600\text{K}$. At both temperature $U_{2}$ and $U_{\rho}$ have been parametrized at $T=550\text{K}$. }
\label{fig:TPBi-transferability}
\end{figure*}

We also have looked into the structural transferability at multiple temperatures. Figure~\ref{fig:TPBi-transferability} shows the radial distribution functions of the A-A, A-B and B-B pairs from both mapped AA and CG simulation at $T=500\,\text{K}$ and $T=600\,\text{K}$, using a CG model that was parameterized at $T=550\,\text{K}$.

\section{Additional details for TCTA model}
\begin{figure*}[ht]
\centering
\includegraphics[width=1.0\linewidth]{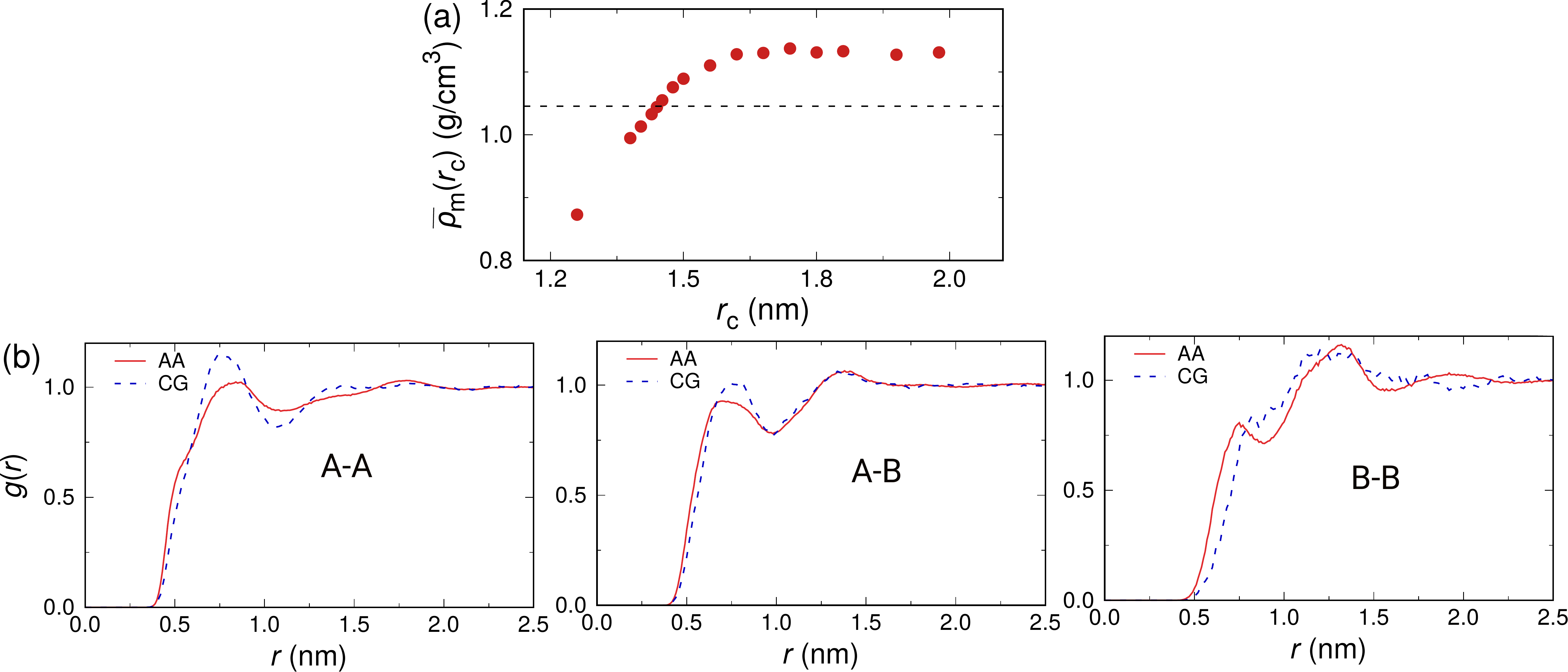}
\caption{(a) Average mass density $\bar{\rho}_\text{m}^\text{CG}$ as a function of cutoff radius $r_\text{c}$ from CG simulations of TCTA. The dashed line indicates $\bar{\rho}_\text{m}^\text{AA}$. (b) Radial distribution functions, $g(r)$, from the mapped AA simulations of TCTA.}
\label{fig:TCTA}
\end{figure*}

We mapped TCTA molecules to four-sites using two particle types, where the LD potential was only used for the A$|$A pairs. Figure~\ref{fig:TCTA}(a) shows the radial distribution functions of the A-A, A-B and B-B pairs from the mapped AA simulations, while Fig.~\ref{fig:TCTA}(b) shows the average mass density as a function cutoff radius $r_\text{c}$.

\section{Additional details for Tm3PyPB model}
\begin{figure*}[ht]
\centering
\includegraphics[width=1.0\linewidth]{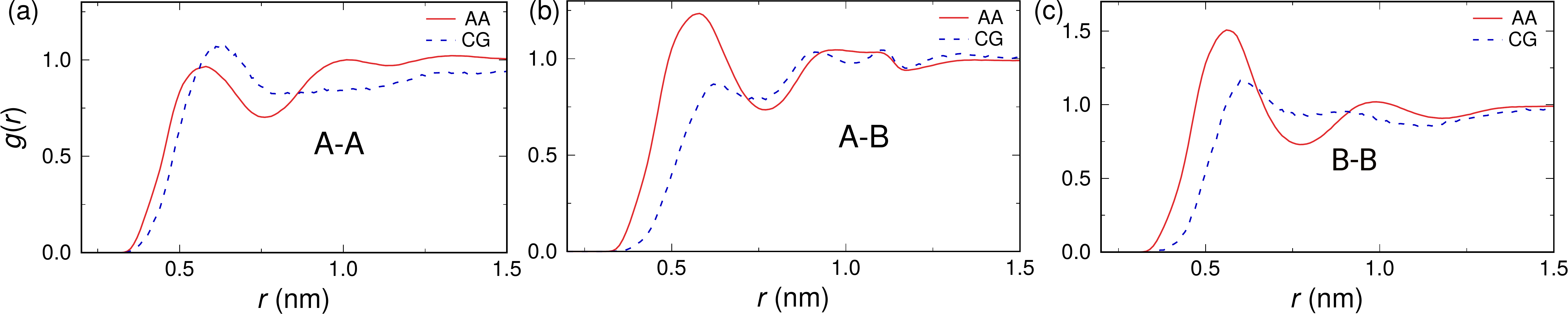}
\caption{(b) Radial distribution functions, $g(r)$, from the CG and AA simulations of Tm3PyPB, with LDP included only for the A$|$A pairs. Simulations performed with $r_\text{c} = 1.5\,\text{nm}$ at $T=550K$.}
\label{fig:Tm3PyPB}
\end{figure*}

Here, we have first included the LDP only for the A$|$A pairs, and systematically scanned $r_\text{c}$. Table~\ref{tab:Tm3PyPB} shows the average mass density ($\bar{\rho}_\text{m}^\text{CG}$) and compressibility $\kappa^\text{CG}$, which deviate strongly from the values obtained from the AA simulations, $\bar{\rho}_\text{m}^\text{AA} = 1.005\,\text{g/cm}^3$ and $\kappa^\text{AA}= 29.13 \times 10^{-5}\,\text{bar}^{-1}$. We also have compared the radial distribution functions, $g(r)$, obtained for $r_\text{c} = 1.5\,\text{nm}$ (Fig.~\ref{fig:Tm3PyPB}). It is evident, that other than A-A, all other pair structures have large deviations. 

\begin{table}[ht]
    \centering
    \begin{tabular}{lcccc}
    \hline
    $r_\text{c}$ (nm) & $\bar{\rho}_\text{m}^\text{CG}$ (g/cm$^3$) & $\kappa^\text{CG}$ ($\times 10^{-5}\,\text{bar}^{-1}$)\\
    \hline
     1.46& 0.48 & 667.85 \\
     1.48& 0.54 & 503.34 \\
     1.50& 0.56 & 340.37 \\
    \hline
    \end{tabular}
    \caption{Average mass density $\bar{\rho}_\text{m}^\text{CG}$ and isothermal compressibility $\kappa^\text{CG}$ from CG simulations at $T=550\,\text{K}$ for Tm3PyPB. LDP used only for A$|$A pairs.} 
    \label{tab:Tm3PyPB}
\end{table}

To improve the CG model, we have added LD contributions to the A$|$B and B$|$A pairs. Since the $g(r)$ for the different particle pairs show features at similar distances (Fig.~\ref{fig:Tm3PyPB}), we scanned the $r_\text{c}$ for the A$|$B and B$|$A pairs exactly as $r_\text{c}$ of A$|$A. The resulting mass density and compressibility are much closer to the AA reference values (Table~\ref{tab:Tm3PyPB2}).

\begin{table}[ht]
    \centering
    \begin{tabular}{lcccc}
    \hline
    $r_\text{c}$ (nm) & $\bar{\rho}_\text{m}^\text{CG}$ (g/cm$^3$) & $\kappa^\text{CG}$ ($\times 10^{-5}\,\text{bar}^{-1}$)\\
    \hline
     1.20& 0.842 & 22.68\\
     1.40& 0.896 & 15.07\\
     1.50& 0.938 & 10.14 \\
     1.55& 0.882 & 9.987\\
     1.60& 0.726 & 13.86 \\
    \hline
    \end{tabular}
    \caption{Average mass density $\bar{\rho}_\text{m}^\text{CG}$ and isothermal compressibility $\kappa^\text{CG}$ from CG simulations at $T=550\,\text{K}$ for Tm3PyPB. LDP used for A$|$A, A$|$B and B$|$A pairs using the same cutoff radius $r_\text{c}$.}
    \label{tab:Tm3PyPB2}
\end{table}

\begin{figure*}[ht]
\centering
\includegraphics[width=1.0\linewidth]{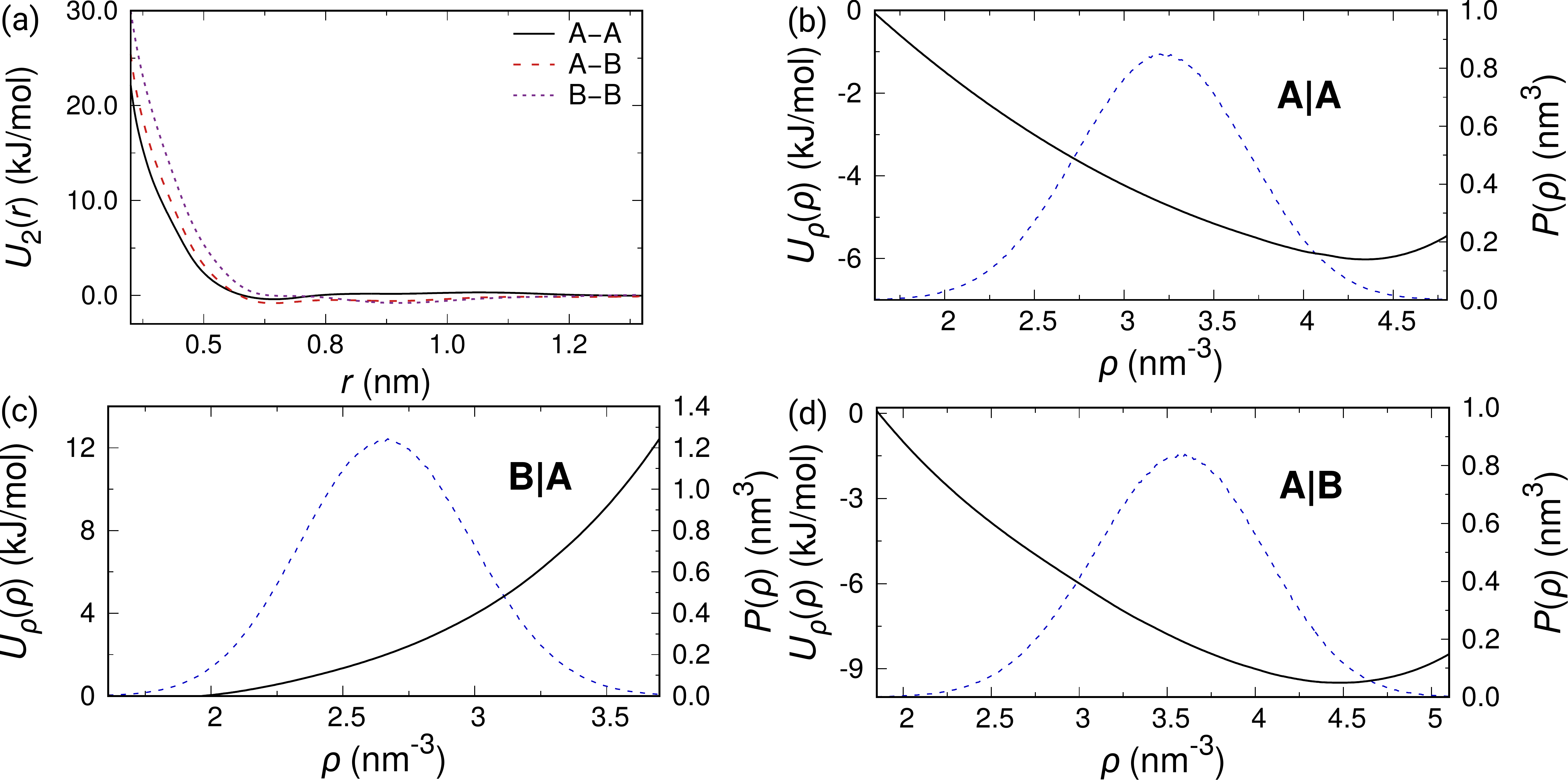}
\caption{(a) Comparison of $U_{2}^{\text{pair-only}}$ for different bead pairs parameterized at $T=550\,\text{K}$. LD potential $U_{\rho}$, parameterized at $T=550\,\text{K}$ for (b) A$|$A, (c) B$|$A, and (d) A$|$B pairs.}
\label{fig:Tm3PyPB2-U2}
\end{figure*}

\begin{figure*}[ht]
\centering
\includegraphics[width=1.0\linewidth]{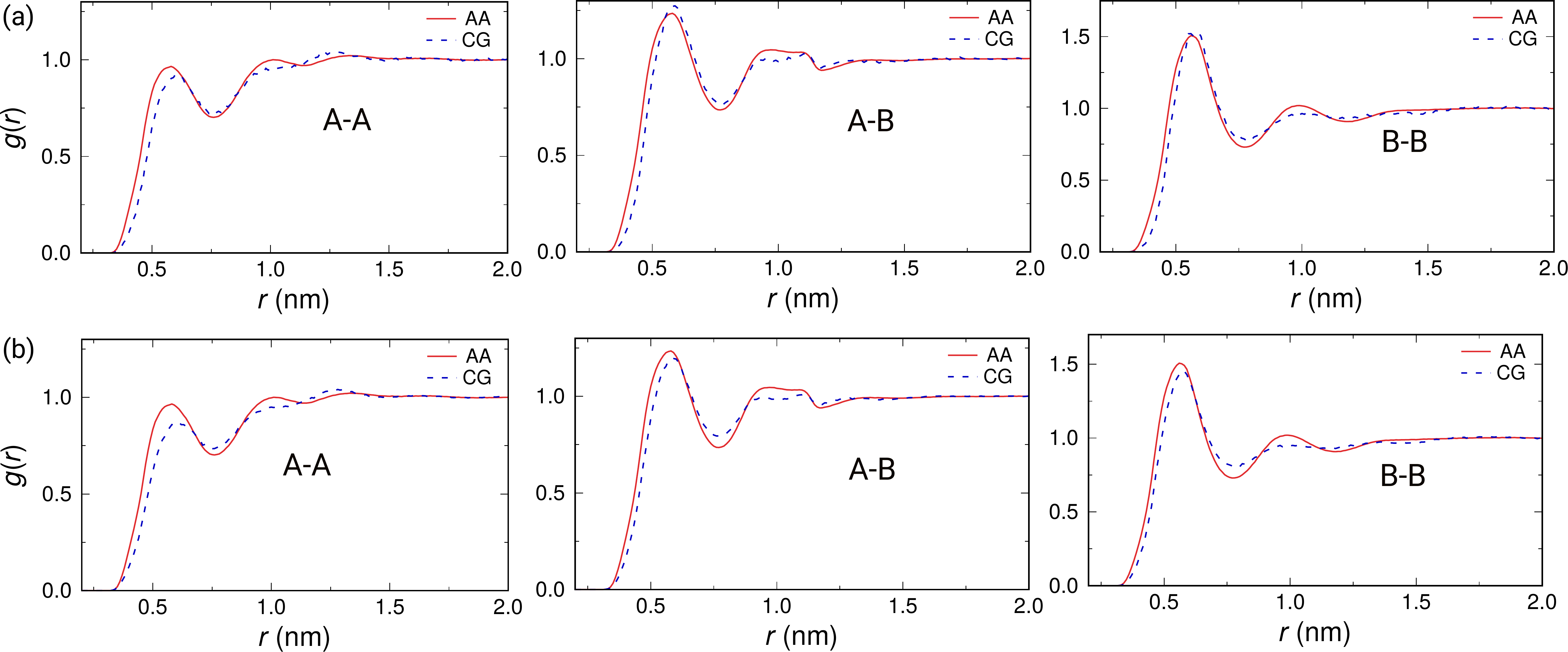}
\caption{Comparison of radial distribution function from AA and CG mapped ensemble (a) at $500 \text{K}$ (b) at 600 \text{K} with $U_2$, $U_{\rho}$ parameterized at $550 \text{K}$.}
\label{fig:Tm3PyPB2-full}
\end{figure*}

\section{Liquid-vapor coexistence}
\begin{figure*}[ht]
\centering
\includegraphics[width=1.0\linewidth]{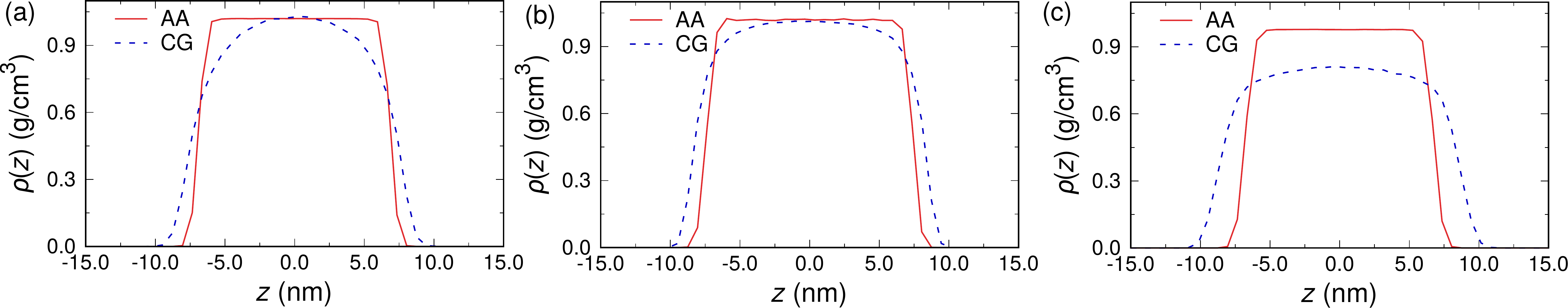}
\caption{Density profiles $\rho_\text{m}(z)$ of (a) mCBP (two-site model), (b) TCTA, and (c) Tm3PyPB from CG simulations, with $U_2^\text{pair-only}$ parameterized from the reference AA slab simulations. All data for $T=550\,\text{K}$.}
\label{fig:slabs-U2}
\end{figure*}

\begin{figure*}[ht]
\centering
\includegraphics[width=1.0\linewidth]{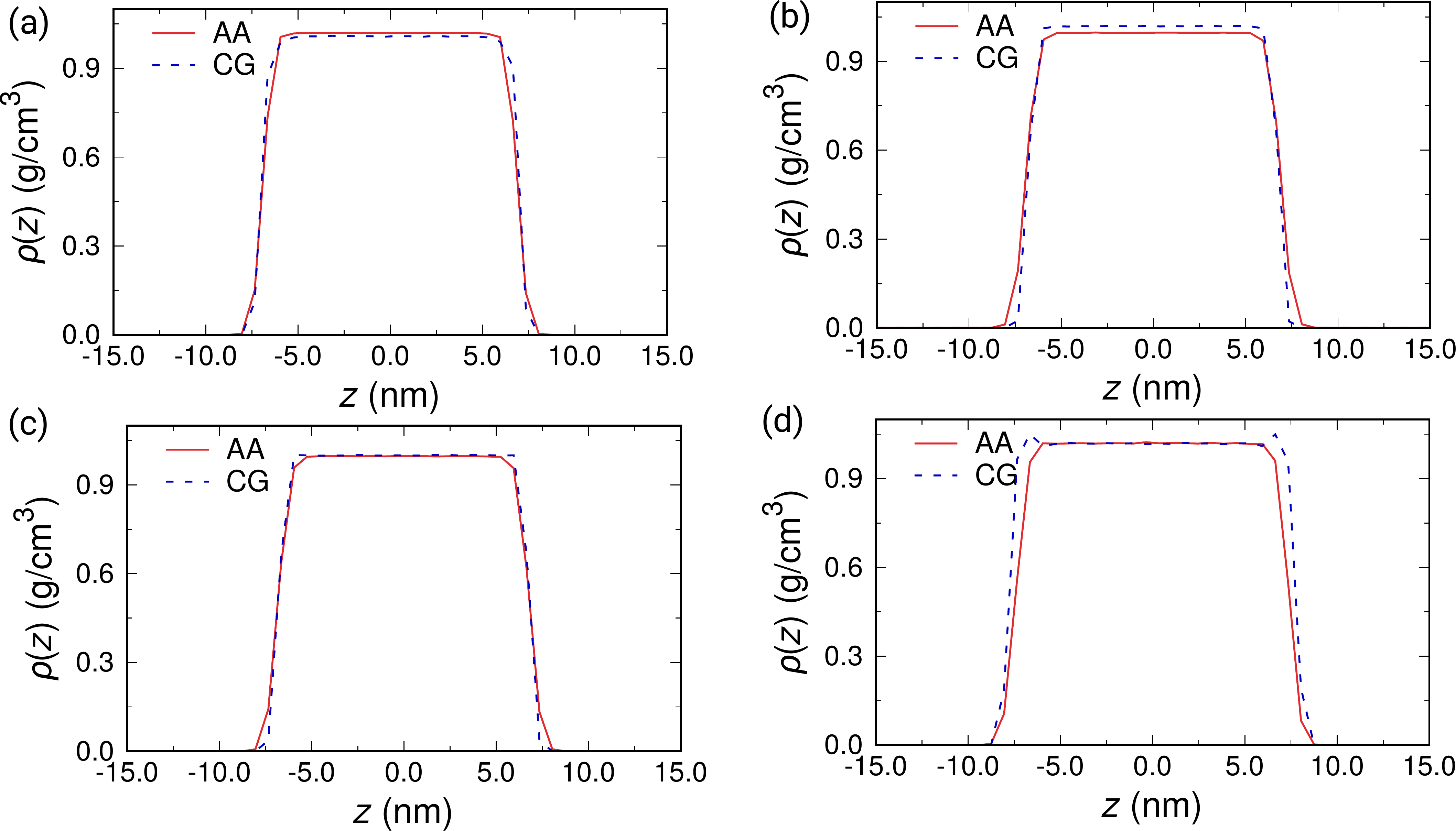}
\caption{Density profiles $\rho_\text{m}(z)$ of (a) mCBP (eight-site model), (b) CBP, (c) BCzPh, and (d) TCTA. All data for $T=550\,\text{K}$.}
\label{fig:slabs2}
\end{figure*}

We have discussed in Sec.~\ref{sec:U2-paironly} the failure of $U_2^\text{pair-only}$ in reproducing stable thin films when parameterizing from bulk AA simulations. When $U_2^\text{pair-only}$ is instead parameterized from reference AA slab simulations under constant $NVT$, the resulting films are stable but the interfaces are much too broad and the coexistence densities deviate from their target (Fig.~\ref{fig:slabs-U2}). In contrast, CG simulations with LDPs reproduce the thin films much more accurately, as shown in Fig.~\ref{fig:slabs2}.

\section{Orientation profile from thin films}
\begin{figure*}[htbp]
\centering
\includegraphics[width=0.6\linewidth]{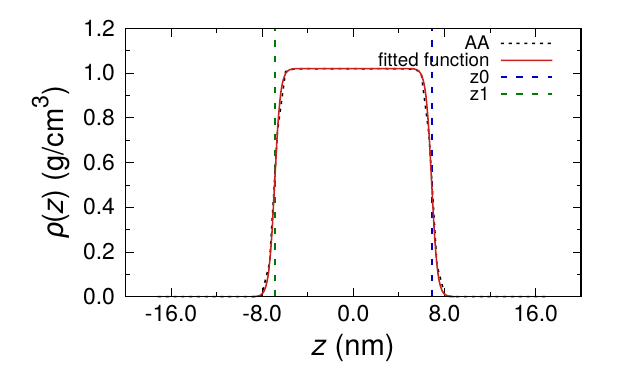}
\caption{Double-tanh fit for the liquid vapor coexistence of mCBP from AA simulations. The vertical lines at $z_0$ and $z_1$ indicate the interface centers.}
\label{fig:tanh}
\end{figure*}

\begin{figure*}[htbp]
\centering
\includegraphics[width=1.0\linewidth]{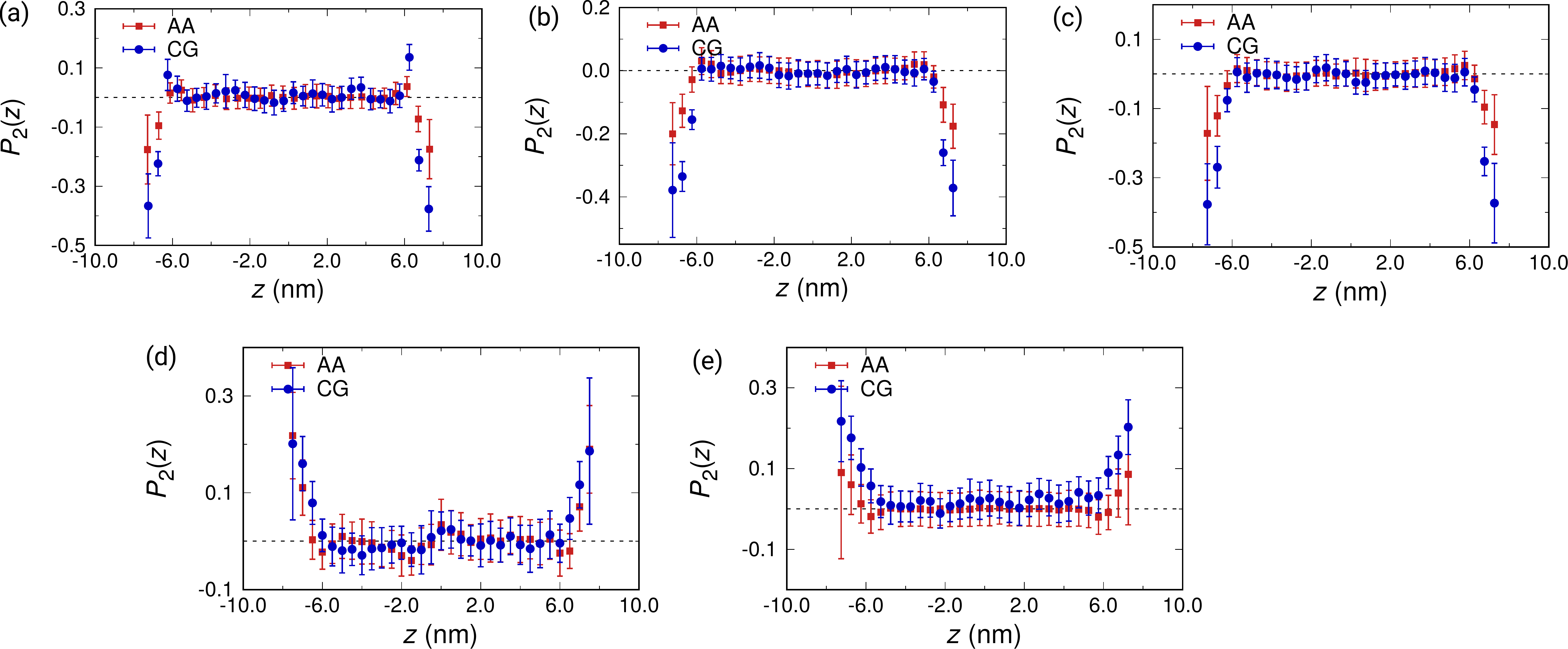}
\caption{Comparison of $P_2(z)$ along the $z$ direction for (a) mCBP (eight-site model), (b) CBP, (c) BCzPh, (d) TCTA, and (e) Tm3PyPB. All data for $T=550\,\text{K}$.}
\label{fig:P2}
\end{figure*}

To define the positions and widths of the liquid--vapor interfaces, we analyzed the density profile $\rho(z)$ along the film normal. The density changes smoothly from the bulk liquid plateau to the low-density region. For the thin films with two opposing interfaces, the following "double--tanh" model was fitted to the density profiles (Fig.~\ref{fig:tanh}):
\begin{equation}
\rho(z) \;=\; \rho_\text{avg} \;+\; \frac{\Delta \rho}{2}
\Bigg[ \tanh\!\left(\frac{z - z_0}{d}\right)
       - \tanh\!\left(\frac{z - z_1}{d}\right) \Bigg],
\end{equation}
where $\rho_\text{avg}$ is the average of the bulk and vapor densities and $\Delta \rho$ is their difference. The parameters $z_0$ and $z_1$ denote the centers of the two interfaces, respectively, and $d$ characterizes the interface width. We have measured the average orientational order parameter, quantified as $\bar{P_2}$, by averaging $P_2(z)$ within the interfacial boundaries $z_{1,2} \pm d$ (Fig.~\ref{fig:P2}). 

\clearpage
\providecommand{\latin}[1]{#1}
\makeatletter
\providecommand{\doi}
  {\begingroup\let\do\@makeother\dospecials
  \catcode`\{=1 \catcode`\}=2 \doi@aux}
\providecommand{\doi@aux}[1]{\endgroup\texttt{#1}}
\makeatother
\providecommand*\mcitethebibliography{\thebibliography}
\csname @ifundefined\endcsname{endmcitethebibliography}
  {\let\endmcitethebibliography\endthebibliography}{}